\documentclass[%
 aip,
 amsmath,amssymb,
 reprint,%
]{revtex4-2}

\usepackage{graphicx}
\usepackage{dcolumn}
\usepackage{bm}

\usepackage[utf8]{inputenc}
\usepackage[T1]{fontenc}
\usepackage{mathptmx}
\usepackage{xcolor}
\usepackage{textcomp}

\DeclareUnicodeCharacter{2606}{"HEREEEE"}

\begin{document}

\preprint{AIP/123-QED}

\title{LayerPCM: An implicit scheme for dielectric screening from
layered substrates}

\author{Jannis Krumland}
 \affiliation{Physics Department and IRIS Adlershof, Humboldt-Universität zu Berlin, 12489 Berlin, Germany}
 
\author{Gabriel Gil}%
\affiliation{Dipartimento di Scienze Chimiche, Universit\`a degli studi di Padova, via F. Marzolo 1, 35131 Padova, Italia}
\affiliation{Instituto de Cibernetica, Matem\'atica y F\'isica, Calle E esq 15 Vedado 10400, La Habana, Cuba}
\email{gabriel@icimaf.cu}

\author{Stefano Corni}
\affiliation{Dipartimento di Scienze Chimiche, Universit\`a degli studi di Padova, via F. Marzolo 1, 35131 Padova, Italia}
\affiliation{CNR—Istituto Nanoscienze, via Campi 213/A, 41125 Modena, Italia}

\author{Caterina Cocchi}
\affiliation{Physics Department and IRIS Adlershof, Humboldt-Universität zu Berlin, 12489 Berlin, Germany}
\affiliation{Institute of Physics, Carl von Ossietzky Universität Oldenburg, 26129 Oldenburg, Germany}
\email{caterina.cocchi@uni-oldenburg.de}

\date{\today}

\begin{abstract}
We present LayerPCM, an extension of the polarizable-continuum model coupled to real-time time-dependent density-functional theory for an efficient and accurate description of the electrostatic interactions between molecules and multilayered dielectric substrates on which they are physisorbed. The former are modelled quantum-mechanically, while the latter are treated as polarizable continua characterized by their dielectric constants. 
The proposed approach is purposely designed to simulate complex hybrid heterostructures, with nano-engineered substrates including a stack of anisotropic layers. 
LayerPCM is suitable to describe the polarization-induced renormalization of frontier energy levels of the adsorbates in the static regime. 
Moreover, it can be reliably applied to simulating laser-induced ultrafast dynamics of the molecules through the inclusion of electric fields generated by Fresnel-reflection at the substrate. 
Depending on the complexity of the underlying layer structure, such reflected fields can assume non-trivial shapes and profoundly affect the dynamics  of the photo-excited charge carriers in the molecule. In particular, the interaction with the substrate can give rise to strong delayed fields which lead to interference effects resembling those of multi-pulse-based spectroscopy.
The robustness of the implementation and the above-mentioned features are demonstrated with a number of examples, ranging from intuitive models to realistic systems.
\end{abstract}
\maketitle
\section{Introduction}
Hybrid materials composed of organic molecules deposited on inorganic substrates have been intensively investigated in the last decade due to their potential for optoelectronic applications.~\cite{Kubatkin2003,Moth-Poulsen2009,nagata2013apl, feng2014jmca, schulz2014afm} Depending on the nature of the constituents, different types of chemical bonds can be formed at the interface.~\cite{brau+09am} Covalent interactions, corresponding to molecular chemisorption, which are typically obtained in the presence of anchor groups, give rise to large degrees of interfacial charge transfer and to pronounced electronic hybridization.~\cite{calz+12jpcc,hofmann2013jcp,lang+14afm,timp+2014cm,turk+19ats,matt+20jpcc} The resulting hybrid system exhibits new characteristics compared to those of its individual building blocks. On the other hand, when molecules are \textit{physisorbed} on a surface, they preserve their intrinsic features to a large extent. Nonetheless, their electronic and optical properties undergo significant variations, including band-gap renormalization~\cite{garc+09prb,della+11prl,pusc+12prb,desp+13prb,nabo+19cm} and reduction of exciton binding energies.~\cite{desp+13prb,fu+17pccp,kerf+18jcp} 

Capturing these effects with the required accuracy represents one of the main challenges for modern electronic structure theory.~\cite{draxl2014acr} Many-body perturbation theory (MBPT), including the $GW$ approximation and the solution of the Bethe-Salpeter equation, is the state-of-the-art first-principles method to compute quasi-particle band structures and optical absorption spectra including correlated electron-hole pairs (\textit{i.e.}, excitons).~\cite{onid+02rmp} Unfortunately, the computational costs of these calculations remain prohibitive for realistic materials described by hundreds of atoms, in spite of the remarkable efforts devoted lately to scale up the performance of MBPT calculations.~\cite{intr1,intr2,intr3,intr4,intr5} Therefore, the \textit{ab initio} treatment of complex systems including nanostructured substrates and/or mesoscopic elements, such as nanoparticles and nanotips, remains unfeasible with purely atomistic schemes. 

A viable workaround for the aforementioned scenarios is provided by implicit schemes that capture \textit{effectively} the electrostatic interactions between photo-active systems and their environment.
Such methods have been applied for decades in computational chemistry to describe molecules in solution. The polarizable-continuum model (PCM)~\cite{tomasi2005cr} and the conductor-like screening model~\cite{cosmo} are successful examples of approaches that have been specifically tailored to deal with isotropic (liquid) media surrounding an active solute. More recently, embedding schemes coupled to MBPT calculations have been proposed and successfully applied~\cite{intr10,duch+18cs,wehn+18jctc,tiri+20jcp,kshi+20jctc} even in the context of periodic systems.~\cite{intr11,intr12,intr13}
However, these advances were devised to deal with molecules in solution or on bulk substrates, but not with substrates consisting of multiple stacked confined layers.

The advent of two-dimensional (2D) materials~\cite{novo+04sci,sple+10nl} and their heterostructures,~\cite{geim-grig13nat} which are known to feature also complex patterns,~\cite{zhan+17sciadv,pan+18nl} have made this scenario very realistic, paving the way for hybrid materials with nano-engineered inorganic substrates.~\cite{intr6,intr7,intr8,intr9} 
Suitable theoretical methods that are able to treat this complexity at affordable computational costs are therefore needed for substantial advances in the field.

In this work, we present an extension of the PCM coupled to real-time time-dependent density-functional theory (RT-TDDFT)\cite{liang+2012jpca, nguyen+2012jpcl, pipolo+2014ctc} for an efficient and yet accurate description of the electrostatic interactions between quantum-mechanically modelled physisorbed molecules and implicit dielectric substrates. Envisioning the application of this method to stacked 2D materials, we allow the substrate to feature a layered structure exhibiting in-plane/out-of-plane anisotropy. We demonstrate that the developed formalism can be employed to calculate the polarization-induced renormalization of frontier energy levels of the adsorbate. For simulations of laser-induced ultrafast dynamics with RT-TDDFT, we additionally supply a way of calculating electric fields that arise from Fresnel-reflection at the substrate. We show that, depending on the complexity of the layered substrate, these reflected fields can assume non-trivial shapes and profoundly affect the electron dynamics induced in the molecule. As an example, we consider a substrate containing a deep-lying dielectric mirror, which gives rise to a delayed field after irradiation with a laser pulse. This leads to interference effects in the coherent dynamics of the molecule, similar to those of multi-pulse-based spectroscopy.
We finally compare our numerical results with those obtained from an exactly solvable two-level model, in order to disclose the dynamics of the excited electrons and the role of polarization effects therein.

This paper is organized as follows. In Section~\ref{pcm.sec}, we provide a brief review of the non-equilibrium flavor of the PCM in the integral-equation formalism, laying out the essential equations. On this basis, we present our extension of the model in Section~\ref{deriv.sec}, together with the proposed method of calculating the Fresnel-reflected fields. In Section~\ref{impl.sec}, we review how the PCM is interfaced with TDDFT, and provide computational details. Finally, in Section~\ref{results.sec}, we report numerical results that verify our implementation, and showcase different types of substrate-related effects that can be described with the presented method. Conclusions are given in Section~\ref{conclu.sec}.

\section{Background: The polarizable-continuum model}\label{pcm.sec}
\begin{figure}
    \centering
    \includegraphics[width=0.4\textwidth]{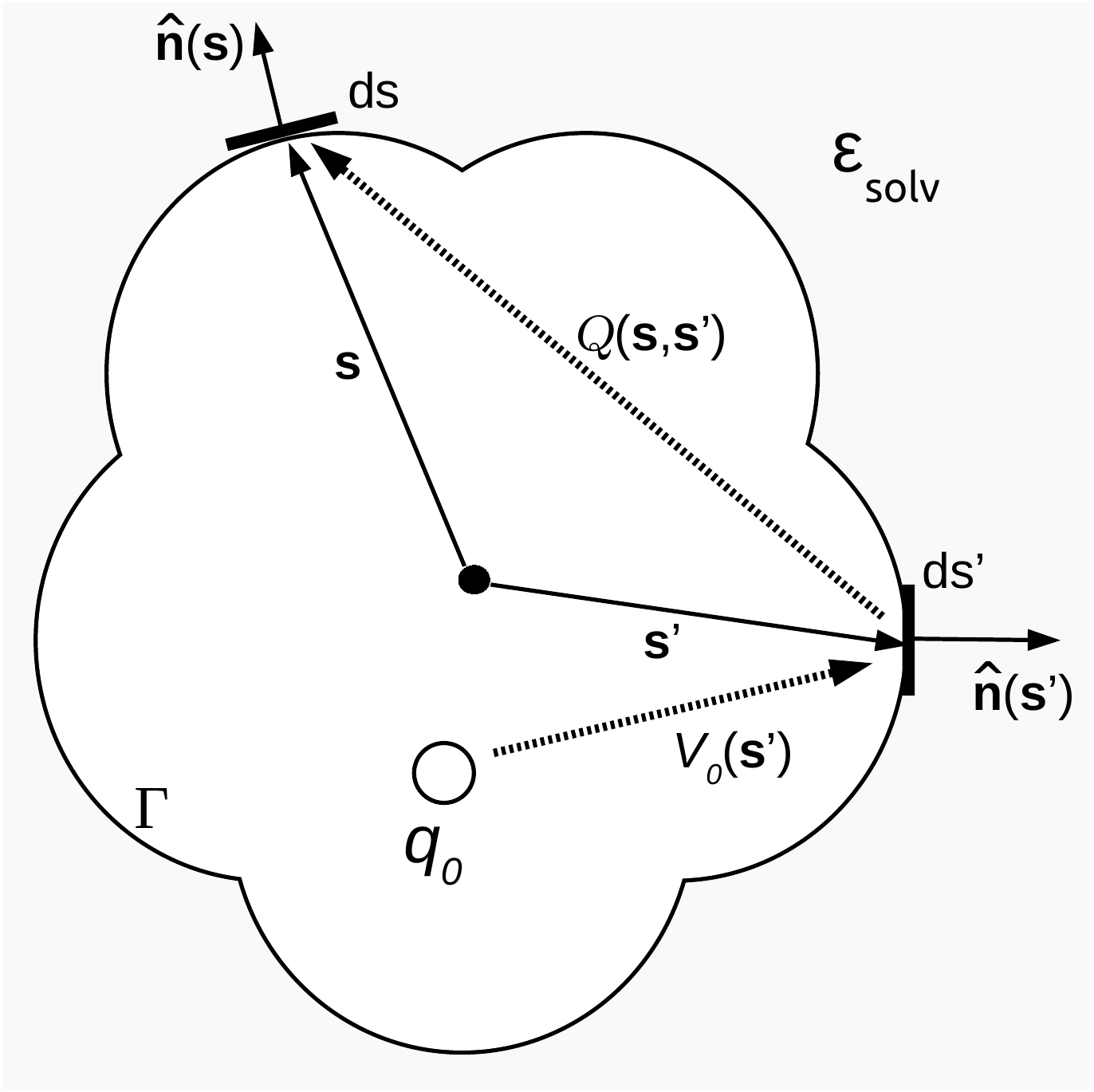}
    \caption{Sketch of the cavity, indicating the quantities that enter in the calculation of the PCM response function. $q_0$ stands for a free charge within the cavity that gives rise to the electrostatic potential $V_0$.}
    \label{fig.sketch}
\end{figure}

The PCM poses a well-established way of coupling quantum-chemically modelled molecules to macroscopic polarizable environments.~\cite{tomasi2005cr} It was originally conceived to study the influence of electrostatic solute-solvent interactions on the properties of solvated systems.~\cite{onsager1936jacs, miertus+1981} In this framework, the solute molecule resides in a molecularly shaped cavity cut into an infinitely extended dielectric medium, which is typically characterized by its dielectric constant (see Fig.~\ref{fig.sketch}). The nuclear and electronic charge densities of the molecule polarize the surrounding medium, giving rise to polarization surface charge density $\sigma_0$ spread on the cavity boundary~$\Gamma$. These polarization charges cause an electrostatic field, the so-called \textit{reaction field}, that, in turn, polarizes the solute inside the cavity. The Schr{\"o}dinger equation for the solute within the reaction field and the Poisson equation for the molecular charge densities coupled to the continuum dielectric are solved self-consistently. 

For a given molecular charge density that generates the electrostatic potential $V_0$, the polarization surface charge density is calculated as
\begin{align}\sigma_0(\textbf{s}) = \int_{\Gamma}\text ds'\,{\cal Q}_0(\textbf{s},\textbf{s}' )V_0(\textbf{s}'),
\label{asc_static.eq}\end{align}
where ${\cal Q}_0$ is the PCM response function, connecting the potential at the cavity point $\textbf{s}'$ with a polarization-induced apparent surface charge $\sigma_0(\textbf{s})\text ds$ at the cavity point $\textbf{s}$ (see Fig.~\ref{fig.sketch}). The combined effect of all these charges gives rise to the electrostatic reaction potential 
\begin{align}\label{reaction_pot.eq}
    V_{\text{R}}(\textbf{r}) = \int_\Gamma \frac{\sigma_0(\textbf{s})\text ds}{|\textbf{r}-\textbf{s}|}.
\end{align}
Applying Eqs.~\eqref{asc_static.eq} and \eqref{reaction_pot.eq} corresponds to solving Poisson's equation within the cavity with a "free" charge density given by the molecular electronic and nuclear charges, correctly taking into account all boundary conditions at the dielectric interfaces. As the next step, the charge density of the molecule subject to the field of Eq.~\eqref{reaction_pot.eq} is recalculated, and the new electrostatic potential, $V_0$, is inserted back into Eq.~\eqref{asc_static.eq}. This procedure is repeated until self-consistency is achieved.

Within the integral-equation formalism,~\cite{cances1997jcp, cances1998jmc} the response function ${\cal Q}_0$, which is the crucial quantity in the PCM, is calculated from the static dielectric constant of the solvent, $\varepsilon$, via
\begin{align}
{\cal Q}_0 = &{\cal Q}(\varepsilon) = \left\lbrace\left[2\pi I-D_e(\varepsilon)\right]S_i+S_e(\varepsilon)\left(2\pi I+D_i^*\right)\right\rbrace^{-1}\nonumber\\&\times\left\lbrace S_e(\varepsilon)S_i^{-1}\left(2\pi I+D_i\right)+\left[2\pi I+D_e(\varepsilon)\right]\right\rbrace\label{pcmmat.eq}
\end{align}
where $I$ is the identity operator, and $S_{i}$, $D_{i}$ and $D^*_i$, as well as $S_{e}$ and $D_{e}$ are integral operators, defined by their action on an function $u$ supported on the cavity surface $\Gamma$ as\cite{cances1998jmc}
\begin{subequations}\label{eq.internal_matrices}
\begin{align}
    \left(S_{i}u\right)(\textbf{s}) &= \int_\Gamma\text ds'\, G_{i}(\textbf{s},\textbf{s}')u(\textbf{s}') \label{simat.eq}\\
    \left(D_{i}u\right)(\textbf{s}) &= \int_\Gamma\text ds'\left[\hat{\textbf{n}}(\textbf{s}')\cdot \nabla' G_{i}(\textbf{s},\textbf{s}')\right]u(\textbf{s}') \label{dimat.eq}\\
    \left(D^*_iu\right)(\textbf{s}) &= \int_\Gamma\text ds'\left[\hat{\textbf{n}}(\textbf{s})\cdot \nabla G_{i}(\textbf{s},\textbf{s}')\right]u(\textbf{s}')\label{dimatstar.eq},
\end{align}
with the electrostatic Green's function for vacuum, $G_i(\textbf{r}, \textbf{r}')=1/|\textbf{r}-\textbf{r}'|$, and
\end{subequations}
\begin{subequations}\label{eq.external_matrices}
\begin{align}
    \left[S_{e}(\varepsilon)u\right](\textbf{s}) &= \int_\Gamma\text ds'\, G_{e}(\textbf{s},\textbf{s}';\varepsilon)u(\textbf{s}') \label{semat.eq}\\
    \left[D_{e}(\varepsilon)u\right](\textbf{s}) &= \varepsilon_{\mathrm{solv}}\int_\Gamma\text ds'\left[\hat{\textbf{n}}(\textbf{s}')\cdot \nabla' G_{e}(\textbf{s},\textbf{s}';\varepsilon)\right]u(\textbf{s}').\label{demat.eq}
\end{align}
\end{subequations}
Here, $\hat{\textbf{n}}(\textbf{s})$ is the normal vector at the surface point $\textbf{s}$, and $\varepsilon_{\mathrm{solv}}$ is the dielectric constant of the medium encompassing the cavity (see Fig.~\ref{fig.sketch}). $G_e$, appearing in Eqs.~\eqref{eq.external_matrices}, solves Poisson's equation for a free point charge in the environment,  disregarding the response of the cavity surface itself. The application of $S_x$ and $D_x$ operators to a function $u$ thus yields the electrostatic potential of a polarization surface charge density $u$ and a surface dipole density $u\hat{\textbf{n}}$, respectively, \textit{in vacuo} ($x=i$) or in the environment ($x=e$). For an isotropic solvent with dielectric constant $\varepsilon_{\mathrm{solv}}$, $G_e(\textbf{r}, \textbf{r}';\varepsilon)=G_e^0(\textbf{r}, \textbf{r}';\varepsilon_{\mathrm{solv}})=1/\varepsilon_{\mathrm{solv}} |\textbf{r}-\textbf{r}'|$. In the generalized case, $\varepsilon$ represents all dielectric constants characterizing the environment. 

While the formalism is based on electrostatics, it can be adopted for time-dependent calculations as well. The short distances between solute and surrounding solvent molecules justify a quasi-static treatment of the dynamical electromagnetic interactions, \textit{i.e.}, the neglect of retardation. In this case, however, it is not very accurate to characterize the solvent by its dielectric constant alone. Often, a non-equilibrium version~\cite{marcus1956jcp, aguilar+1993jcp, cammi+tomasi1995ijqc, menucci+jcp1998} of time-dependent PCM is employed, which additionally includes the optical dielectric constant of the solvent, $\varepsilon_d$. Time scales are separated: The inert rotational degrees of freedom of the solvent are associated with the static dielectric constant, whereas the fast polarization response of the electrons in the solvent is described by $\varepsilon_d$. In ultrafast dynamical processes such as vertical ionizations or optical excitations, this non-equilibrium approach assumes that the electronic degrees of freedom are able to instantaneously follow the rapidly varying electrostatic forces generated by the charge densities of the solute, while the rotational (and in general the nuclear) solvent degrees of freedom remain fully frozen. The time-dependent apparent surface charge density is then given by~\cite{aguilar+1993jcp, cammi+tomasi1995ijqc, menucci+jcp1998, ingrosso+2003jml, caricato+2005jcp}
\begin{align}\sigma(\textbf{s},t) &= \sigma_0(\textbf{s})+\int_{\Gamma}\text ds'\,{\cal Q}_d(\textbf{s},\textbf{s}') [V(\textbf{s}',t)-V_0(\textbf{s}')],
\label{asc.eq}\end{align}
where $V(t)$ is the instantaneous electrostatic potential caused by the molecule in its dynamical perturbed state, and ${\cal Q}_d$ is the dynamical response function, calculated as ${\cal Q}_d$~=~${\cal Q}(\varepsilon_d)$ with the same prescription as in Eq.~\eqref{pcmmat.eq}. Eq.~\eqref{asc.eq} is valid as long as the solvent has no optical excitations in the considered frequency range. For a more general treatment, one should account for the frequency-dependent and complex-valued nature of the dielectric function of the solvent,~\cite{ding2015jcp, corni2015jpca, pipoleCorni2016jpcc} which, however, is not straightforward and goes beyond the scope of the present work.

While originally devised for molecules immersed in isotropic solvents, the reformulation of PCM within the integral-equation formalism\cite{cances1997jcp, cances1998jmc} allows for the overhead-free featuring of more general dielectric environments, such as anisotropic\cite{cances1997jcp} and ionic\cite{mennucci+1997jpcb, cances1998jmc} solvents, metal surfaces,~\cite{corni2003jcp} close-by nanoparticles,~\cite{corni2001jcp, delgado2013jcp} and interfaces between different fluid phases.~\cite{frediani2004jcp,frediani2004jpcb} 
The Green's function for the interior of the cavity, $G_i$, remains unchanged, whereas the one for the environment, $G_e$, assumes a different form. In Section~\ref{deriv.sec}, we determine $G_e$ for a stack of anisotropic dielectric layers.



\section{Electromagnetic response of the substrate}\label{deriv.sec}

\begin{figure}
    \centering
    \includegraphics[width=0.47\textwidth]{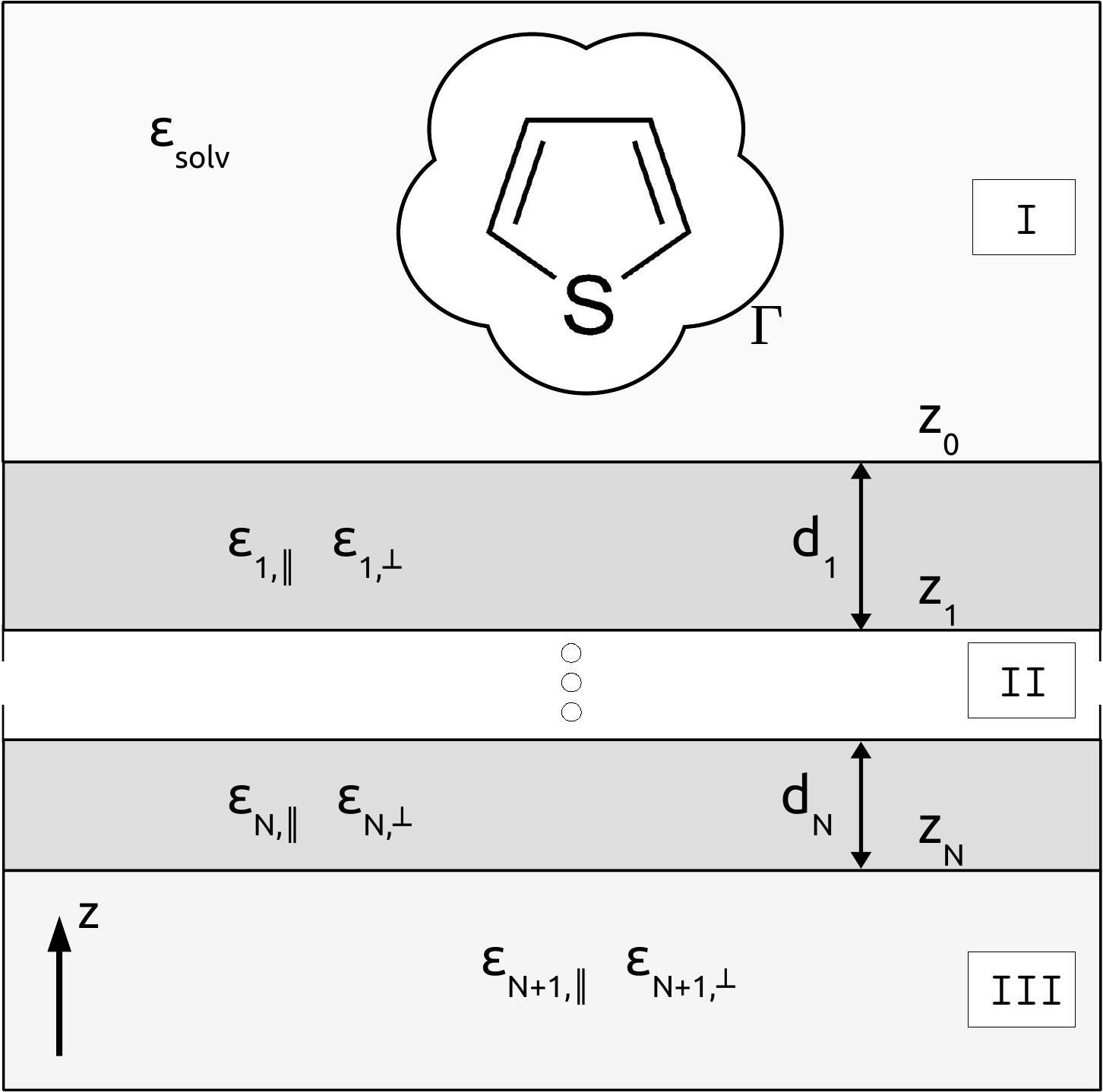}
    \caption{Sketch of the dielectric environment of a thiophene molecule residing in a cavity with surface $\Gamma$. The molecule is surrounded by a solvent with dielectric constant $\varepsilon_\mathrm{solv}$. The substrate is formed by $N$ layers of thickness $d_n$ between the coordinates $z=z_{n-1}$ and $z=z_{n}$, and has in-plane and out-of-plane dielectric constants $\varepsilon_{n,\parallel}$ and $\varepsilon_{n,\perp}$, respectively. The final layer ($N+1$) extends down to $z=-\infty$.}
    \label{env.fig}
\end{figure}

In this section, we derive the equations for the description of molecule-substrate-solvent interactions (Section~\ref{green.sec}) and of the reflection behavior of an electromagnetic plane wave impinging on the substrate (Section~\ref{fresnel.sec}). The whole environment of the thiophene ring considered in this example is built up as sketched in Fig.~\ref{env.fig}. The molecule is immersed in a solvent with dielectric constant $\varepsilon_{\mathrm{solv}}$, residing in a vacuum cavity with boundary $\Gamma$ (region I). The vacuum cavity represents the solvent-inaccessible surrounding of the thiophene ring.~\cite{tomasi2005cr} The layered substrate is characterized by the origin and normal vector of its surface, by the perpendicular and parallel components of the dielectric constant, $\varepsilon_{n,\perp}$ and $\varepsilon_{n,\parallel}$, respectively, as well as by the thicknesses $d_n$ of the $N$ intermediate layers (region II). The semi-infinite bottom layer is characterized by dielectric constants $\varepsilon_{N+1,\parallel}$ and $\varepsilon_{N+1,\perp}$ (region III). In the following, we refer to the parallel direction ($\parallel$) as ``in-plane", and to the perpendicular one ($\perp$) as ``out-of-plane".

\subsection{The Green's function of the environment}\label{green.sec}
We make use of the integral-equation formalism of the PCM to calculate the reaction field stemming from the polarization of the substrate and the solvent. The missing ingredient is the electrostatic Green's function of the environment, $G_e$, to be inserted into Eqs.~\eqref{eq.external_matrices}, which depends parametrically on the sets of dielectric constants $\{\varepsilon\}$, on the layer thicknesses $\{d\}$, and on the shape and position of the cavity. To keep the notation simple, we omit these dependencies in the following. The dynamical PCM response function, ${\cal Q}_d$, to be plugged into Eq.~\eqref{asc.eq} for non-equilibrium time-dependent calculations, is obtained by replacing $\{\varepsilon\}$ by the set of optical dielectric constants, $\{\varepsilon_d\}$.

The environmental Green function, $G_e$~=~$G_e(\textbf{r},\textbf{r}')$, is the total electrostatic potential at a point~$\textbf{r}$ generated by a unit point charge situated at~$\textbf{r}'$, ignoring the electrostatic response of the vacuum cavity. From Eqs.~\eqref{reaction_pot.eq}, \eqref{pcmmat.eq}, and \eqref{eq.external_matrices}, it is evident that this function is ultimately evaluated for $\textbf{r}$ and $\textbf{r}'$ lying on the cavity surface $\Gamma$; thus, both these vectors can be assumed to be within region I (see Fig.~\ref{env.fig}). The electrostatic potential in this region is the sum of the direct solvent-screened Coulomb potential and of a second term stemming from the polarization of the substrate, $G_{\text{img}}$, such that:
\begin{align}
G_e(\textbf{r}, \textbf{r}') = G_e^0(\textbf{r}, \textbf{r}')+G_{\text{img}}(\textbf{r},\textbf{r}').\label{solv.eq}
\end{align}
By definition, this function solves Poisson's equation, which, restricting ourselves to the solvent region, reads
\begin{align}
    -\varepsilon_{\mathrm{solv}}\nabla^2G_e(\textbf{r}, \textbf{r}')= \delta(\textbf{r}-\textbf{r}').\label{poisson.eq}
\end{align}
From Eqs. \eqref{solv.eq} and \eqref{poisson.eq}, it is clear that the polarization term satisfies Laplace's equation, 
\begin{align}
    -\nabla^2G_{\mathrm{img}}(\textbf{r}, \textbf{r}')= 0.\label{laplaceSolv.eq}
\end{align}

Extending previous work within the PCM,~\cite{corni2003jcp, frediani2004jcp,frediani2004jpcb} the following derivations exploit the cylindrical symmetry of the dielectric environment, which implies that $G_e(\textbf{r}, \textbf{r}')$ depends on the in-plane variables $x$, $x'$, $y$, and $y'$ only via $\Delta r_\parallel$ =  $|\Delta\textbf{r}_\parallel|$, where $\Delta\textbf{r}_\parallel= (x-x',y-y')$.
    Applying a partial spatial Fourier transform $\textbf{r}_\parallel\rightarrow\textbf{q}_\parallel$, we can write\cite{corni2002jcp}
\begin{align}
    G_e(\textbf{r},\textbf{r}') = \frac{1}{2\pi}\int_0^\infty\text dq_\parallel\,q_\parallel J_0(q_\parallel\Delta r_\parallel)G(q_\parallel, z, z'),\label{greenPartial.eq}
\end{align}
where $J_0$ is the zeroth-order Bessel function of first kind.~\cite{temme1996} Due to in-plane isotropy, the transformed Green's function, $G(q_\parallel, z, z')$, depends only on the absolute value~$q_\parallel$ of the in-plane wave vector~$\textbf{q}_\parallel$. 
By transformation of Eq.~\eqref{solv.eq}, we obtain\cite{fuchs1981prb}
\begin{align}
G_e(q_\parallel, z, z') = \frac{2\pi}{\varepsilon_{\mathrm{solv}} q_\parallel}e^{-q_\parallel|z-z'|}+G_{\text{img}}(q_\parallel, z, z'),
\end{align}
where $G_{\text{img}}(q_\parallel, z, z')$ satisfies the transformed version of Eq.~\eqref{laplaceSolv.eq},
\begin{align}
\left(-q_\parallel^2+\frac{\partial^2}{\partial z^2}\right)G_{\text{img}}(q_\parallel, z, z') = 0.\label{greenSolvent.eq}
\end{align}
Eq.~\eqref{greenSolvent.eq} is evidently solved by
\begin{align}
    G_{\text{img}}(q_\parallel, z, z') = B_0(q_\parallel,z')\,e^{-q_\parallel(z-z_0)},\label{imageFourier.eq}
\end{align}
where the exponentially increasing solution has been discarded, as it diverges for $z\rightarrow\infty$. To simplify later expressions, the zero of $z$ is set to the surface at $z_0$. All electrostatic effects due to the polarization of a specific substrate are contained in the constant of integration, $B_0(q_\parallel,z')$. We obtain the latter by enforcing appropriate boundary conditions at the surface and at all underlying interfaces in the substrate. To this end, we need to know the electrostatic potential inside the layers. With the assumed in-plane/out-of-plane anisotropy, the dielectic tensor in layer $n$ in region II is given by $\varepsilon_n = \text{diag}\{\varepsilon_{n,\parallel},\varepsilon_{n,\parallel},\varepsilon_{n,\perp}\}$, and the electrostatic potential satisfies a generalized Laplace's equation,
\begin{align}
     \left(-\varepsilon_{n,\parallel} q_\parallel^2+\varepsilon_{n,\perp}\frac{\partial^2}{\partial z^2}\right)G_{\text{e},n}(q_\parallel,z,z') &= 0,\label{anisoLaplace.eq}
     \end{align}
     which is a direct consequence of Gauss' law, $\nabla\cdot\textbf{D} = \nabla\cdot(\varepsilon\textbf{E})= 0$. Eq. \eqref{anisoLaplace.eq} is solved by
     \begin{align}
    G_{\text{e},n}(q_\parallel,z,z') &= A_ne^{\xi_n q_\parallel (z-z_{n-1})}+ B_ne^{-\xi_n q_\parallel (z-z_{n-1})}, \label{greenSubstrate.eq}
\end{align}
where $\xi_n$~=~$(\varepsilon_{n,\parallel}/\varepsilon_{n,\perp})^{1/2}$ is a measure of the in-plane/out-of-plane anisotropy, and $z_{n-1}$ marks the interface in the direction of the surface.

The next step towards calculating $B_0(q_\parallel,z')$ [see Eq.~\eqref{imageFourier.eq}] consists of connecting the potentials in Eq.~\eqref{greenSubstrate.eq} at the interfaces inside the substrate according to the boundary conditions $\delta\textbf{E}_\parallel$~=~$0$ and $\delta\textbf{D}_\perp$~=~$0$, which are derived from the basic electrostatic equations $\nabla\times\textbf{E}$~=~$0$ and $\nabla\cdot\textbf{D}$~=~$0$, respectively. Since its curl vanishes, the electric field $\textbf{E}$ is the gradient of the potential~$G_\text{e}$, and the boundary conditions are equivalent to $\delta G_\text{e}(q_\parallel, z, z')$~=~$0$ and $\delta [\varepsilon_\perp\partial G_\text{e}(q_\parallel, z, z')/\partial z]$~=~$0$, where the $\delta$-variation is related to the $z$~coordinate. We consider the interface between the layers $n$ and $n+1$, which is located at $z_n<z_0$. Thus, the continuity condition, $\delta G(q_\parallel, z, z') = 0$, corresponds to
\begin{subequations}\label{eq.cont_and_deriv}
\begin{align}
A_ne^{-\xi_nq_\parallel d_n}+ B_ne^{\xi_n q_\parallel d_n} = A_{n+1}+ B_{n+1},\label{cont}
\end{align}
where $d_n = z_{n-1}-z_n$ is the thickness of the layer $n$. The second condition,
$\delta [\varepsilon_\perp\partial G(q_\parallel, z, z')/\partial z] = 0$, reads
\begin{align}
\xi_n\varepsilon_{n,\perp}\left(A_ne^{-\xi_nq_\parallel d_n}- B_ne^{\xi_n q_\parallel d_n}\right) = \xi_{n+1}\varepsilon_{n+1,\perp}\left(A_{n+1}- B_{n+1}\right).\label{deriv}
\end{align}
\end{subequations}
Eqs. \eqref{eq.cont_and_deriv} constitute a linear system that can be rewritten in matrix form as
 \begin{align}
 \begin{pmatrix}
A_n\\B_n
 \end{pmatrix}
 =
 {\cal D}_n(q_\parallel)
 \begin{pmatrix}
A_{n+1}\\B_{n+1}
 \end{pmatrix}, 
 \end{align}
 where 
 \begin{widetext}
 \begin{align}
 {\cal D}_n(q_\parallel) =
     \frac{1}{2\xi_n\varepsilon_{n,\perp}}\begin{pmatrix}
(\varepsilon_{n,\perp}\xi_n+\varepsilon_{n+1,\perp}\xi_{n+1})e^{q_\parallel\xi_n d_n} & 
 (\varepsilon_{n,\perp}\xi_n-\varepsilon_{n+1,\perp}\xi_{n+1})\,e^{q_\parallel\xi_n d_n}
 \\
(\varepsilon_{n,\perp}\xi_n-\varepsilon_{n+1,\perp}\xi_{n+1})\,e^{-q_\parallel\xi_n d_n} & 
(\varepsilon_{n,\perp}\xi_n+\varepsilon_{n+1,\perp}\xi_{n+1})e^{-q_\parallel\xi_n d_n}
 \end{pmatrix}.
\end{align}
\end{widetext}
The matrix ${\cal D}_n$ connects the potentials in the $n$'th and $(n+1)$'th layers. The coefficients in the first layer, $n$~=~1, and in the final semi-infinite layer, $n$~=~$N+1$, are thus connected via
\begin{align}
 \begin{pmatrix}
A_1\\B_1
 \end{pmatrix}
 =
 {\cal T}(q_\parallel)
 \begin{pmatrix}
A_{N+1}\\0
 \end{pmatrix}, \label{transferMatrixProp.eq}
 \end{align}
where the \textit{transfer matrix} is defined as
\begin{align}
 {\cal T}(q_\parallel) = \prod_{n=1}^{N}{\cal D}_n(q_\parallel).
\end{align}
In Eq.~\eqref{transferMatrixProp.eq}, we have made explicit that the coefficient $B_{N+1}$ vanishes in region III, since it corresponds to an unphysical exponentially rising solution for $z\rightarrow-\infty$. 

Finally, we consider the boundary conditions at the surface, \textit{i.e.} at the interface between regions I and II, where $z=z_0$. From Eqs. \eqref{imageFourier.eq} and \eqref{greenSubstrate.eq} for $n=1$, we have
\begin{subequations}\label{eq.surface_boundary}
\begin{align}
    \frac{2\pi}{\varepsilon_{\mathrm{solv}}q_\parallel}e^{q_\parallel(z_0-z')}+B_0 &= A_1 + B_1\label{surfaceBoundary1.eq}\\
    2\pi e^{q_\parallel(z_0-z')}-{\varepsilon_0}q_\parallel B_0 &= q_\parallel\varepsilon_1^\perp\xi_1(A_1 - B_1).\label{surfaceBoundary2.eq}
\end{align}
\end{subequations}
One of the constants, $A_1$ or $B_1$, can be eliminated by using Eq.~\eqref{transferMatrixProp.eq}, which implies $A_1/B_1 = {\cal T}_{11}/{\cal T}_{21}$. Consequently, Eqs. \eqref{eq.surface_boundary} form a linear system with two equations for two unknowns ($B_0$ and $A_1$ or $B_1$), which can be solved to yield the desired constant $B_0$, and thus an expression for the polarization term of the Green's function:
\begin{align}
G_{\text{img}}(q_\parallel,z,z') = \frac{2\pi}{\varepsilon_{\mathrm{solv}}q_\parallel}\frac{\varepsilon_{\mathrm{solv}}-\varepsilon_{\mathrm{subs}}(q_\parallel)}{\varepsilon_{\mathrm{solv}}+\varepsilon_{\mathrm{subs}}(q_\parallel)}e^{-q_\parallel(z'+z-2z_0)},\label{bZero.eq}
\end{align}
where $\varepsilon_{\mathrm{subs}}(q_\parallel) = \varepsilon_{1,\perp}\xi_1[\,{\cal T}_{11}(q_\parallel)-{\cal T}_{21}(q_\parallel)\,]/[\,{\cal T}_{11}(q_\parallel)+{\cal T}_{21}(q_\parallel)\,]$.

In order to build $D_e$ according to Eq.~\eqref{demat.eq}, the $\textbf{r}'$-gradient of the Green's function is required. With the polarization term in the form of Eq.~\eqref{greenPartial.eq} and the relations
\begin{subequations}
\begin{align}
    \frac{\partial G_{\text{img}}(q_\parallel,z,z')}{\partial z'} &= -q_\parallel G_{\text{img}}(q_\parallel,z,z'),\\
    \frac{\text dJ_0}{\text dx} &= -J_1(x),\\
    \text{and}\hspace{0.5cm}\nabla'(\Delta r_\parallel) &= -\frac{\Delta\textbf{r}_\parallel}{\Delta r_\parallel} = -\Delta\hat{\textbf{r}}_\parallel,
\end{align}
\end{subequations}
where $J_1$ is the first-order Bessel function of first kind,~\cite{temme1996} we obtain\cite{corni2002jcp}
\begin{align}
 \nabla' G_{\text{img}}(\textbf{r},\textbf{r}')
    &=-\frac{1}{2\pi}\int_0^\infty\text dq_\parallel\,q_\parallel^2 G_{\text{img}}(q_\parallel, z, z')\nonumber\\
    &\times\left[ J_0(q_\parallel\Delta r_\parallel)\hat{\mathbf{z}}-J_1(q_\parallel\Delta r_\parallel){\Delta\hat{\textbf{r}}_\parallel}\right].\label{nablaGreenPartial.eq}
\end{align}
Eqs.~\eqref{bZero.eq} and \eqref{nablaGreenPartial.eq} are used to calculate the electrostatic response functions ${\cal Q}_0$ and ${\cal Q}_d$.

\subsection{Fresnel reflection}\label{fresnel.sec}
\subsubsection{Reflectivity}
The electrons and nuclei polarize the substrate, causing an electric field that acts back on the molecule. If an external time-dependent electric field, \textit{e.g.} associated with electromagnetic radiation, acts on the system, the corresponding charge densities change in time, leading to a time-dependent reaction field. However, the external field interacts also directly with the layered substrate. At each interface, it is partially reflected, and the reflected beams interfere with the incident one. To determine the field acting locally on the molecule, these effects should be accounted for. In order to calculate the reflected fields, we make use once more of the transfer matrix formalism to calculate the frequency-dependent reflectivity of the substrate. This is common practice for anisotropic dielectric materials with arbitrary numbers of layers.~\cite{yeh1979josa, yeh1980ss, puschnig2006aem, vorwerk2016cpc, passler2017josab, passler2020prb} Full anisotropy requires dealing with 4$\times$4 matrices, as it mixes $s$- and $p$-modes at the interfaces between different dielectrics. In the present case of mere in-plane/out-of-plane anisotropy (uniaxial birefringence), the symmetry of the problem is still high enough to inhibit this type of mixing. Consequently, the 4$\times$4 matrix equations collapse into separate 2$\times$2 systems for $s$- and $p$-polarizations. Thus, we can avoid cumbersome numerical procedures and solve the problem analytically. Finally, with the knowledge of the frequency-dependent reflectivity, we can compute the total field as a superposition of the incident wave and of all partially reflected waves in real time.

The reaction field is calculated in the quasistatic limiting case, which amounts to the neglect of retardation. This approximation is known to be accurate at short distances, but cannot be applied to the external field: unlike the short-ranged field emitted by the molecule, the external, incident field is a plane wave, which is not attenuated over distance. Hence, it interacts also with deep substrate layers, whereas the substrate-molecule coupling is restricted to sub-nanometric depths within the surface. Hence, the full set of Maxwell's equations has to be considered. As the application of the transfer-matrix formalism to calculate reflectivities is well-established,~\cite{yeh1979josa, yeh1980ss, puschnig2006aem, vorwerk2016cpc, passler2017josab, passler2020prb} we present here only a reduced derivation of the formulas and refer the reader to the Appendix for further details.

The differential Maxwell's equations within a layer with in-plane/out-of-plane anisotropy can be cast into a Fourier-space wave equation
\begin{align}\label{wave.eq}
k^2\textbf{E}+\frac{\varepsilon_\perp-\varepsilon_\parallel}{\varepsilon_\parallel}\textbf{k}k_zE_z + \left(\frac{\omega}{c}\right)^2\varepsilon\textbf{E}= 0.
\end{align}
The condition for nontrivial solutions of the linear system in Eq.~\eqref{wave.eq} gives rise to a number of allowed modes, characterized by their wave vector $\textbf{k}$; the corresponding solutions $\textbf{E}$ determine the associated polarization directions of the field. We assume that $\textbf{k}=\textbf{k}_\mathrm{in}$ is known outside of the substrate. Its in-plane component, $\textbf{k}_\parallel$, remains unchanged throughout all layers. The determinant is a polynomial of 6th order in $k_z$ and, thus, has six independent roots. However, Gauss' law, $\nabla\cdot\textbf{D} = \nabla\cdot(\varepsilon\textbf{E}) = 0$, imposes additional constraints, resulting in only four possible wave vectors:
\begin{subequations}
\begin{align}
\textbf{k}^{(s)}_\pm &= \left(k_\parallel,0, \pm\sqrt{\left(\frac{\omega}{c}\right)^2\varepsilon_\parallel-k_\parallel^2}\right) = \left(k_\parallel,0, \pm k_z^{(s)}\right),\label{kS} \\
\textbf{k}^{(p)}_\pm &= \left(k_\parallel, 0, \pm\sqrt{\left(\frac{\omega}{c}\right)^2\varepsilon_\parallel-\frac{\varepsilon_\parallel}{\varepsilon_\perp}k_\parallel^2}\right) = \left(k_\parallel,0, \pm k_z^{(p)}\right)\label{kP},
\end{align}
\end{subequations}
where the coordinate system is rotated in the $x$-$y$ plane such that $\textbf{k}_\parallel$ points to the $x$ direction. We also introduce for convenience the $z$-direction wavenumbers, $k_z^{(s)}$ and $k_z^{(p)}$. The corresponding polarization vectors are
\begin{subequations}
\begin{align}
{\boldsymbol{\gamma}}^{(s)}_\pm &= \left(0, 1, 0\right)\label{pS}\\
{\boldsymbol{\gamma}}^{(p)}_\pm &\propto \left(1, 0, \mp\frac{\varepsilon_\parallel}{\varepsilon_\perp}\frac{k_\parallel}{k_z^{(p)}}\right).\label{pP}
\end{align}
\end{subequations}
    The modes characterized by the wave vectors in Eq.~\eqref{kS} and by the polarizations in Eq.~\eqref{pS} are forward (+) and backward (-) propagating $s$-polarized waves, with polarization perpendicular to the plane of incidence. As the corresponding field is exclusively in-plane, these modes are not affected by the anisotropy. Consequently, they are purely transversal modes, \textit{i.e.}, $\nabla\cdot\textbf{E}\sim i\textbf{k}^{(s)}\cdot{\boldsymbol{\gamma}}^{(s)} = 0$. The polarization vectors of the $p$-polarized counterparts, Eq. \eqref{kP} and Eq. \eqref{pP}, lie in the plane of incidence. Thus, they have components in both in-plane and out-of-plane directions and are affected by the corresponding anisotropy. As a consequence, $\textbf{k}^{(p)}$ and ${\boldsymbol{\gamma}}^{(p)}$ are not mutually perpendicular. In other words, $p$-polarized waves induce space charge densities $\nabla\cdot\textbf{E}\neq 0$ in anisotropic layers and therefore gain a longitudinal component. 

Since $s$- and $p$-polarizations are decoupled, they can be treated individually. For the $s$-polarization, we start from the electric field. In the Appendix, we show how the reflectivity of the whole layered structure, $r^{(s)}(\omega)$, can be calculated by matching the field amplitudes of adjacent layers at the interfaces, in accordance with the boundary conditions $\delta\textbf{E}_\parallel = 0$ and $\delta\textbf{B}_\parallel = 0$. The final result is
\begin{align}
r^{(s)}(\omega) = {\cal T}^{(s)}_{21}/{\cal T}^{(s)}_{11},\label{rS}
\end{align}
where the transfer matrix related to the electric field is defined as
\begin{align}\label{ts}
    \left.
    {\cal T}^{(s)} = 
    \begin{pmatrix}
    k_{\mathrm{solv},z}^{(s)} & -1 \\
    k_{\mathrm{solv},z}^{(s)} & 1
    \end{pmatrix}
    \right(\prod_{n=1}^{N}
    {\cal D}_n^{(s)}
    \left)
    \begin{pmatrix}
    1 & 1 \\
    -k_{N+1,z}^{(s)} & k_{N+1,z}^{(s)}
    \end{pmatrix},\right.
\end{align}
with 
\begin{align}\label{ds}
{\cal D}_n^{(s)} = 
    \begin{pmatrix}
    \cos(k_{n,z}^{(s)}d_n) & -i\sin(k_{n,z}^{(s)}d_n)/k_{n,z}^{(s)} \\
    -ik_{n,z}^{(s)}\sin(k_{n,z}^{(s)}d_n) & \cos(k_{n,z}^{(s)}d_n)
    \end{pmatrix}.
\end{align}
The case of a $p$-polarized field is more conveniently treated starting from the magnetic field. In this case, we end up with the electric-field reflectivity
\begin{align}
r^{(p)}(\omega) = -{\cal T}^{(p)}_{21}/{\cal T}^{(p)}_{11},\label{rP}
\end{align}
including the magnetic-field-related transfer matrix
\begin{widetext}
\begin{align}\label{tp}
    \left.
    {\cal T}^{(p)} = 
    \begin{pmatrix}
    k_{\mathrm{solv},z}^{(p)}/\varepsilon_{\mathrm{solv}} & -1 \\
    k_{\mathrm{solv},z}^{(p)}/\varepsilon_{\mathrm{solv}} & 1
    \end{pmatrix}
    \right(\prod_{n=1}^{N}
    {\cal D}_n^{(p)}
    \left)
    \begin{pmatrix}
    1 & 1 \\
    -k_{N+1,z}^{(p)}/\varepsilon_{N+1,\parallel} & k_{N+1,z}^{(p)}/\varepsilon_{N+1,\parallel}
    \end{pmatrix},
    \right.
\end{align}
\end{widetext}
where
\begin{align}\label{dp}
{\cal D}_n^{(p)} = 
    \begin{pmatrix}
    \cos(k_{n,z}^{(p)}d_n) & -i(k_{n,z}^{(p)}/\varepsilon_{n,\parallel})^{-1}\sin(k_{n,z}^{(p)}d_n) \\
    -i(k_{n,z}^{(p)}/\varepsilon_{n,\parallel})\sin(k_{n,z}^{(p)}d_n) & \cos(k_{n,z}^{(p)}d_n)
    \end{pmatrix}.
\end{align}
 
We continue by using the determined reflectivity to calculate from an arbitrary incident field the reflected field in time domain.

\subsubsection{Total field close to the surface}
The interaction between the molecule and the external electric field is usually treated in the dipole approximation, since the molecules are typically much smaller ($\sim$1~nm) than the wavelengths of interest (100-1000~nm), and higher multipole components couple to spatial variations of the field rather than to its absolute values. Within this approximation, the molecule is subject to a spatially uniform electric field. We write the incident field as 
\begin{align}
    \textbf{E}_{\text{in}}(t) = \hat{\boldsymbol{\gamma}}E_{\text{in}}(t),\label{incField.eq}
\end{align}
with the complex-valued polarization vector $\hat{\boldsymbol{\gamma}}$, and the real-valued temporal profile $E_{\text{in}}(t)$. From this, the time-dependence of the two components of the reflected field is calculated through causality-respecting convolutions
\begin{subequations}\label{eq.response_both}
\begin{align}
    E_{\text r}^{(s)}(t) &= \int_{-\infty}^t\text dt'\,r^{(s)}(t-t')E_{\text{in}}(t')\label{response_s.eq}\\
    E_{\text r}^{(p)}(t) &= \int_{-\infty}^t\text dt'\,r^{(p)}(t-t')E_{\text{in}}(t'),\label{response_p.eq}
\end{align}
\end{subequations}
where the response functions, $r^{(s)}(t-t')$ and $r^{(p)}(t-t')$, are obtained through inverse Fourier transformations of the reflectivities $r^{(s)}(\omega)$ and $r^{(p)}(\omega)$ [Eqs. \eqref{rS} and \eqref{rP}], respectively. The reflected fields of Eqs.~\eqref{eq.response_both} are weighted by the projection of the polarization vector of the incident field, $\hat{\boldsymbol{\gamma}}$,  onto the $s$- and $p$-polarization vectors. The calculation of these polarization vectors requires the knowledge of the plane of incidence. Even though the spatial dependence of the field is neglected, the wavevector ${\textbf{k}}_{\text{in}}$ of the incident field is needed as an input, since the plane of incidence is spanned by ${\textbf{k}}_{\text{in}}$ and the surface normal vector $\hat{\textbf{n}}$. Since the direction of ${\textbf{k}}_{\text{in}}$ alone is sufficient, and this vector is perpendicular to the polarization $\hat{\boldsymbol{\gamma}}$, this corresponds to specifying a single additional parameter, which is the angle of incidence. We obtain
\begin{subequations}
\begin{align}
    \hat{\boldsymbol{\gamma}}^{(s)}_{\text{in}} &= {\textbf{k}}_{\text{in}}\times\hat{\textbf{n}}/|{\textbf{k}}_{\text{in}}\times\hat{\textbf{n}}|,\label{incidenta.eq}\\
    \hat{\boldsymbol{\gamma}}^{(p)}_{\text{in}} &= {\textbf{k}}_{\text{in}}/|\textbf{k}_{\text{in}}|\times\hat{\boldsymbol{\gamma}}^{(s)}_{\text{in}}.\label{incidentb.eq}
\end{align}
\end{subequations}
The reflected field propagates with the surface-reflected wavevector $\textbf{k}_\text{in}-2\hat{\textbf{n}}\left(\hat{\textbf{n}}\cdot{\textbf{k}}_{\text{in}}\right)$. The associated polarization vectors are
\begin{subequations}
\begin{align}
  \hat{\boldsymbol{\gamma}}^{(s)}_r &= \hat{\boldsymbol{\gamma}}^{(s)}_{\text{in}}\label{reflecteda.eq}\\
    \hat{\boldsymbol{\gamma}}^{(p)}_\text{r} &= -\left[{\textbf{k}}_{\text{in}}-2\hat{\textbf{n}}\left(\hat{\textbf{n}}\cdot{\textbf{k}}_{\text{in}}\right)\right]/|\textbf{k}_{\text{in}}|\times\hat{\boldsymbol{\gamma}}^{(s)}_{\text{in}}.\label{reflectedb.eq}
\end{align}
\end{subequations}
Having defined these quantities, we can write the total field acting on the molecule as 
\begin{align}
    \textbf{E}_{\text{tot}}(t) = \textbf{E}_{\text{in}}(t) + \hat{\boldsymbol{\gamma}}^{(s)}_\text{r}\left(\hat{\boldsymbol{\gamma}}^{(s)}_\text{in}\cdot\hat{\boldsymbol{\gamma}}\right)E_\text{r}^{(s)}(t)+\hat{\boldsymbol{\gamma}}^{(p)}_\text{r}\left(\hat{\boldsymbol{\gamma}}^{(p)}_\text{in}\cdot\hat{\boldsymbol{\gamma}}\right)E_\text{r}^{(p)}(t).
\end{align}
With this expression, it is implicitly assumed that $\textbf{r}$~=~0 for all fields, \textit{i.e.} the field is evaluated directly at the surface. Further away, the reflected field would have an additional delay related to the time it takes to travel from the molecule to the surface and back. This retardation effect, noticeable only for relatively large distances (on the order of visible light wavelengths), can be accounted for by including a corresponding vacuum layer on top of the substrate.

\section{Implementation}\label{impl.sec}
\subsection{Interface with density-functional theory}
The presented formalism can be interfaced easily with density functional theory\cite{hohenberg1964} (DFT) and its time-dependent extension\cite{rungegross} (TDDFT). DFT is a quantum-mechanical framework to determine ground-state electronic structure of a many-electron system. This is achieved by solving the Kohn-Sham equation,\cite{kohnsham1965}
\begin{align}\label{ks.eq}
     \hat{\cal H}_{\text{KS}}[\rho_0]\psi_k(\textbf{r}) = E_k\psi_k(\textbf{r}),
\end{align}
for the Kohn-Sham orbitals $\psi_k(\textbf{r})$ and eigenvalues $E_k$. Eq.~\eqref{ks.eq} must be solved self-consistently, since the Kohn-Sham Hamiltonian depends on the solutions $\psi_k(\textbf{r})$ of the equation via the ground-state electron density
\begin{align}
    \rho_0(\textbf{r}) = \sum_k^{\text{occ}}|\psi_k(\textbf{r})|^2.
\end{align}
The Kohn-Sham Hamiltonian \textit{in vacuo} is 
\begin{align}\label{ks_ham.eq}
    \hat{\cal H}_{\text{KS},0}[\rho_0] = -\frac{\nabla^2}{2}+ V_{\text{nuc}}(\textbf{r})+V_{\text{H}}[\rho_0](\textbf{r})+V_{\text{xc}}[\rho_0](\textbf{r}),
\end{align}
where $V_{\text{xc}}[\rho_0](\textbf{r})$ is the exchange-correlation potential. This term has to be approximated, since its exact form is unknown. $V_{\text{nuc}}(\textbf{r})$ and 
\begin{align}
    V_{\text{H}}[\rho_0](\textbf{r}) = \int\text d^3r'\,\frac{\rho_0(\textbf{r}')}{|\textbf{r}-\textbf{r}'|}
\end{align}
are the electrostatic potentials due to the nuclei and the electron cloud, respectively, up to a sign. They can thus be directly be inserted into Eq.~\eqref{asc_static.eq} via
\begin{align}
    V_0(\textbf{s}') = -\lbrace V_{\mathrm{nuc}}(\textbf{s}')+V_H[\rho_0](\textbf{s}')\rbrace
\end{align}
to obtain $\sigma_0$, \textit{i.e.} the static polarization surface charge density. The polarizing effect of the reaction field on the solute is included by adding a corresponding electrostatic energy to Eq.~\eqref{ks_ham.eq},
\begin{align}\label{total_static_ham.eq}
\hat{\cal H}_{\text{KS}} =
\hat{\cal H}_{\text{KS},0}
-\int_{\Gamma}\text ds\,\frac{\sigma_0(\textbf{s})}{|\textbf{r}-\textbf{s}|}.
\end{align}
Including the polarization of the environment adds another layer of self-consistency to the calculation.

In TDDFT, the time-dependent Kohn-Sham equation
\begin{align}
    i\frac{\partial}{\partial t}\psi_k(\textbf{r},t) = \hat{\cal H}_{\text{KS}}[\rho](t)\psi_k(\textbf{r},t),
\end{align}
is used to propagate in time the orbitals $\psi_k(\textbf{r},t)$ and the dynamical electron density
\begin{align}
    \rho(\textbf{r},t) &= \sum_{k}^{\mathrm{occ}}|\psi_k(\textbf{r},t)|^2,
\end{align}
where the initial condition assumes that the system is in its ground state, \textit{i.e.},  $\psi_k(\textbf{r},t=0)$~=~$\psi_k(\textbf{r})$. In the adiabatic approximation, the full time-dependent Kohn-Sham Hamiltonian including the dynamical reaction field\cite{pipolo+2014ctc,corni2015jpca} is given by 
\begin{align}
    \hat{\cal H}_{\text{KS}}[\rho](t) = \hat{\cal H}_{\mathrm{KS},0}[\rho(t)]-\int_{\Gamma}\text ds\,\frac{\sigma(\textbf{s},t)}{|\textbf{r}-\textbf{s}|}+\textbf{r}\cdot\textbf{E}(t), 
\end{align}
where $\sigma(t)$ is calculated from Eq.~\eqref{asc.eq} with 
\begin{align}
    V(\textbf{s}',t) &= 
    V_H[\rho(t)](\textbf{s}')+V_{\mathrm{nuc}}(\textbf{s}').
\end{align}
The coupling to the external electric field $\textbf{E}(t)$ is included within the dipole approximation. At this stage, we are not including the so-called cavity field effects.~\cite{pipolo+2014ctc, gil2019jctc}

For the analysis of the electron dynamics, we monitor the number of excited electrons,~\cite{krumland2020jcp}
\begin{align}\label{population.eq}
N_e(t) = N - \sum\limits_{i,j}^{\text{occ}}\left|\langle\psi_i(0)|\psi_j(t)\rangle\right|^2,
\end{align}
as well as the time-dependent dipole moment,
\begin{align}\label{dipole.eq}
    \boldsymbol\mu(t) = - \int\text d^3r\, \textbf{r} \, \rho(\textbf{r}, t).
\end{align}
$N_e(t)$ and $\boldsymbol\mu(t)$ give indications about the excited-state populations and coherences of the many-electron system, respectively.~\cite{krumland2020jcp}

\subsection{Computational Details}\label{comp_details.sec}

The presented formalism has been implemented in a module named \textit{LayerPCM} in a locally modified version of the \textsc{Octopus} code (v9.0),~\cite{marques2003cpc, castro2006pssb, andrade2015pccp} incorporating the previous installment of PCM.~\cite{tancogne2020jctc, gil2019jctc, delgado2015jcp} Orbitals are represented on a real-space grid inside a simulation box formed by interlocking spheres centered at the atomic positions. For the first considered system, the He$^+$ ion (Section~\ref{benchmark.sec}), we set the radius to 9.45~bohr; for the second one, the thiophene molecule (1T, Section~\ref{thiophene.sec}), we use 11.34~bohr. We employ grid spacings of 0.28~bohr and 0.36~bohr for He$^+$ and 1T, respectively. For the exchange-correlation potential, we choose the (adiabatic) local-density approximation in the Perdew-Zunger parametrization.~\cite{ekar1984prb, perd1981prb} For $\Delta$self-consistent field ($\Delta$SCF) calculations, we use the corresponding spin-polarized formalism for the 1T$^+$ ion. In time-dependent runs, we employ the approximated enforced time-reversal symmetry propagator and a time step of 0.045~a.u.\cite{cast2004jcp}

The PCM cavities are also build from interlocking spheres.~\cite{pascual1990jcp} In the case of He$^+$, we center one sphere at (0,0,0), which is the position of the ion, and another one at (1~bohr,\,1~bohr,\,1~bohr), with radii of 2.27~bohr each. For 1T, we use a cavity formed by the union of spheres that are centered on the atomic positions; we use radii of 3.78~bohr for C, and 4.54~bohr for S. We do not put spheres on H based on the united-atom approach,~\cite{barone1997jcp} giving rise to a smoother cavity. All these values correspond to the respective van-der-Waals radii of the atoms scaled by a factor 1.2. The cavity surface $\Gamma$ is discretized into $N_t$ surface elements (\textit{tesserae}), characterized by their position $\textbf{s}_k$ and total charge $q_k$. The tessellation transforms the operators in Eqs.~\eqref{eq.internal_matrices} and \eqref{eq.external_matrices}, and thus the PCM response function [Eq.~\eqref{pcmmat.eq}], into square matrices. The reaction potentials become
\begin{align}
    -\int_{\Gamma}\text ds\,\frac{\sigma(\textbf{s})}{|\textbf{r}-\textbf{s}|}\approx -\sum_{k=1}^{N_t}\frac{q_k}{|\textbf{r}-\textbf{s}_k|}.
\end{align}
We employ tessellation densities of 240/sphere and 60/sphere for He$^+$ and 1T, respectively. The total number of tesserae is reduced by discarding those that reside inside the union of the spheres.
The unbounded integrals of Eqs.~\eqref{greenPartial.eq} and \eqref{nablaGreenPartial.eq} are evaluated numerically with a 20-point Gauss-Laguerre quadrature.~\cite{abramowith1974}

\section{Results}\label{results.sec}
In this section, we showcase results that validate the developed approach and the corresponding numerical implementation, including benchmarks against analytical results (Section~\ref{benchmark.sec}). Subsequently, we apply the method to describe a realistic setup in the framework of TDDFT (Section~\ref{thiophene.sec}).

\subsection{He$^+$ ion}\label{benchmark.sec}
To put the developed method and its numerical implementation to the test, we choose the He$^+$ ion as a benchmark system. Due to its spherical symmetry, this cation can be assumed to be reasonably similar to an ideal point charge $+e$. It thus lends itself for comparisons to (semi-)analytical results from elementary electrostatics. As described in Section~\ref{comp_details.sec}, we employ as a cavity the union of two diagonally displaced spheres. Conventionally, a spherical cavity would be used for a single spherical ion; we employ an irregular one, instead, to check the robustness of the implementation also beyond high-symmetry cases. As long as the entire ionic charge density lies within the cavity and the environment is completely on the outside, the shape of the cavity should not matter. 

\begin{figure}
    \centering
    \includegraphics[width=0.48\textwidth]{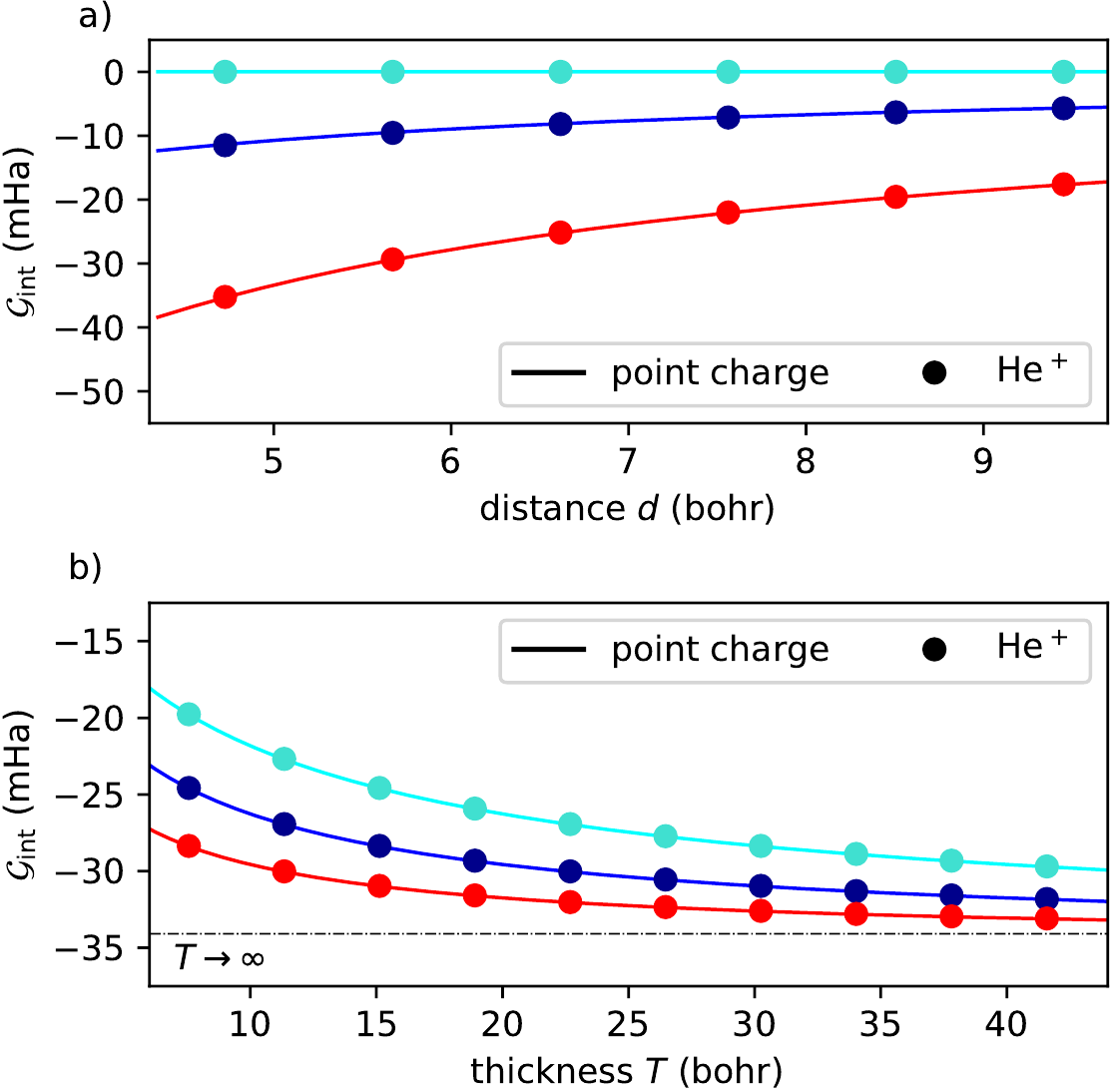}
    \caption{Interaction free energy between a) a charged particle within a solvent with dielectric constant $\varepsilon_{\mathrm{solv}}$ and its induced polarization charges on the surface towards a substrate with dielectric constant $\varepsilon_{\mathrm{subs}}$ at a distance $d$ away [Eq. \eqref{interactionDistance.eq}]. For the turquoise curve, $\varepsilon_{\mathrm{solv}}$~=~$\varepsilon_{\mathrm{subs}}$~=~5; for the blue one, $\varepsilon_{\mathrm{solv}}$~=~2 and $\varepsilon_{\mathrm{subs}}$~=~5; for the red one, $\varepsilon_{\mathrm{solv}}$~=~1 and $\varepsilon_{\mathrm{subs}}$~=~5. b) Interaction free energy between a charged particle (no solvent) and the charges induced in an anisotropic layer of finite thickness $T$ with perpendicular and parallel dielectric constants $\varepsilon_\perp$ and $\varepsilon_\parallel$, respectively [Eq. \eqref{interactionThickness.eq}]. The horizontal dashed line in panel (b) marks the $T\rightarrow\infty$ limit. In both panels, the solid lines are not guides to the eye, but correspond to the analytic electrostatic formulas, whereas the discrete points result from the PCM/DFT formalism, applied to a helium cation (He$^+$).}
    \label{interactionEnergy.fig}
\end{figure}

As a first, textbook example, we consider the $+e$ charge to be within a medium of dielectric constant $\varepsilon_{\mathrm{solv}}$ (which we call \textit{solvent}), placed at a distance $d$ away from the interface to a second semi-infinite medium of dielectric constant $\varepsilon_{\mathrm{subs}}$ (\textit{substrate}). The charge polarizes the solvent and the substrate, giving rise to a reaction field that, in turn, stabilizes the charge. The interaction free energy, defined as the difference of the free energies obtained with and without substrate, is given as
\begin{align}\label{interactionDistance.eq}
{\cal G}_{\text{int}}(d) = -\frac{\varepsilon_{\mathrm{subs}}-\varepsilon_{\mathrm{solv}}}{\varepsilon_{\mathrm{subs}}+\varepsilon_{\mathrm{solv}}}\frac{1}{4\varepsilon_{\mathrm{solv}}d}.
\end{align}
To obtain a comparable quantity from DFT/PCM, we explicitly calculate the free energies in the solvent with and without the substrate, and subtract the resulting values. Looking at the interaction energy as a function of the distance $d$ from the substrate, we find perfect agreement with the analytical expression Eq.~\eqref{interactionDistance.eq} for different combinations of $\varepsilon_{\mathrm{solv}}$ and $\varepsilon_{\mathrm{subs}}$ [see Fig.~\ref{interactionEnergy.fig}a)]. This ideal match has a subtle implication. In the DFT/PCM calculation, the charge is spatially extended and must be hosted by a vacuum cavity. The point charge, on the other hand, does not have such a cavity around it. Thus, there is an additional dielectric interface and also an additional surface charge density in the DFT/PCM scenario. Note that we are referring here to the actual interfacial charge densities, not to the abstract apparent surface charges comprising the reaction of the entire dielectric environment. Interactions between the charges at the cavity boundary and the other polarization-induced charge densities in the substrate should, in principle, give rise to contributions to the interaction energy that are captured by the DFT/PCM calculation, but not by the analytical expression Eq.~\eqref{interactionDistance.eq}. However, the perfect agreement between the two approaches proves that such coupled contributions do not play a role here.

As a second example, we test the description of spatially confined, anisotropic layers. The charge is situated \textit{in vacuo}, at a distance $d$~=~6~bohr away from a single dielectric layer of varying thickness $T$, with perpendicular and parallel dielectric constants $\varepsilon_\perp$ and $\varepsilon_\parallel$, respectively. Since no solvent is included here, it is evident that in this case the cavity and the charges supported on its boundary $\Gamma$ do not bear any physical meaning: $\Gamma$ does not mark an interface between dielectrics, but is situated in free space. Unfortunately, to the best of our knowledge, an analytical benchmark for this dielectric setup does not exist. We therefore use Eqs.~\eqref{greenPartial.eq} and \eqref{bZero.eq} instead. These equations are significantly simplified, since $\Delta r_\parallel$~=~0 and $z'+z-2z_0$~=~$2d$ for a point charge interacting with its own reaction field. We end up with
\begin{align}\label{interactionThickness.eq}
   {\cal G}_{\text{int}}(T) &= -\frac{[(\varepsilon_\perp\varepsilon_\parallel)^{1/2}+1][(\varepsilon_\perp\varepsilon_\parallel)^{1/2}-1]}{2}\times\nonumber\\&\times\int_0^\infty\text dq_\parallel\,\frac{e^{-2q_\parallel d}}{1+\varepsilon_\perp\varepsilon_\parallel+2(\varepsilon_\perp\varepsilon_\parallel)^{1/2}\coth(q_\parallel(\varepsilon_\parallel/\varepsilon_\perp)^{1/2} T)}.
\end{align}
Comparing the interaction energies obtained  from the DFT/PCM calculation and from Eq.~\eqref{interactionThickness.eq} as a function of the layer thickness $T$ and for different anisotropies, we get again essentially identical results [Fig.~\ref{interactionEnergy.fig}b)]. On a side note, Fig. \ref{interactionEnergy.fig}b) highlights the effect on the anisotropy on the reaction field. The product $\varepsilon_\perp\varepsilon_\parallel$ is the same for all three displayed cases, which, according to Eq.~
\eqref{interactionThickness.eq}, means that the curves converge to the same value for $T\rightarrow\infty$. However, the higher the value of the in-plane dielectric constant $\varepsilon_\parallel$ with respect to the out-of-plane one $\varepsilon_\perp$, the more rapidly this limit is approached. Thus, large in-plane dielectric constants give rise to stronger reaction fields than large out-of-plane ones. 

\begin{figure}
    \centering
    \includegraphics[width=.47\textwidth]{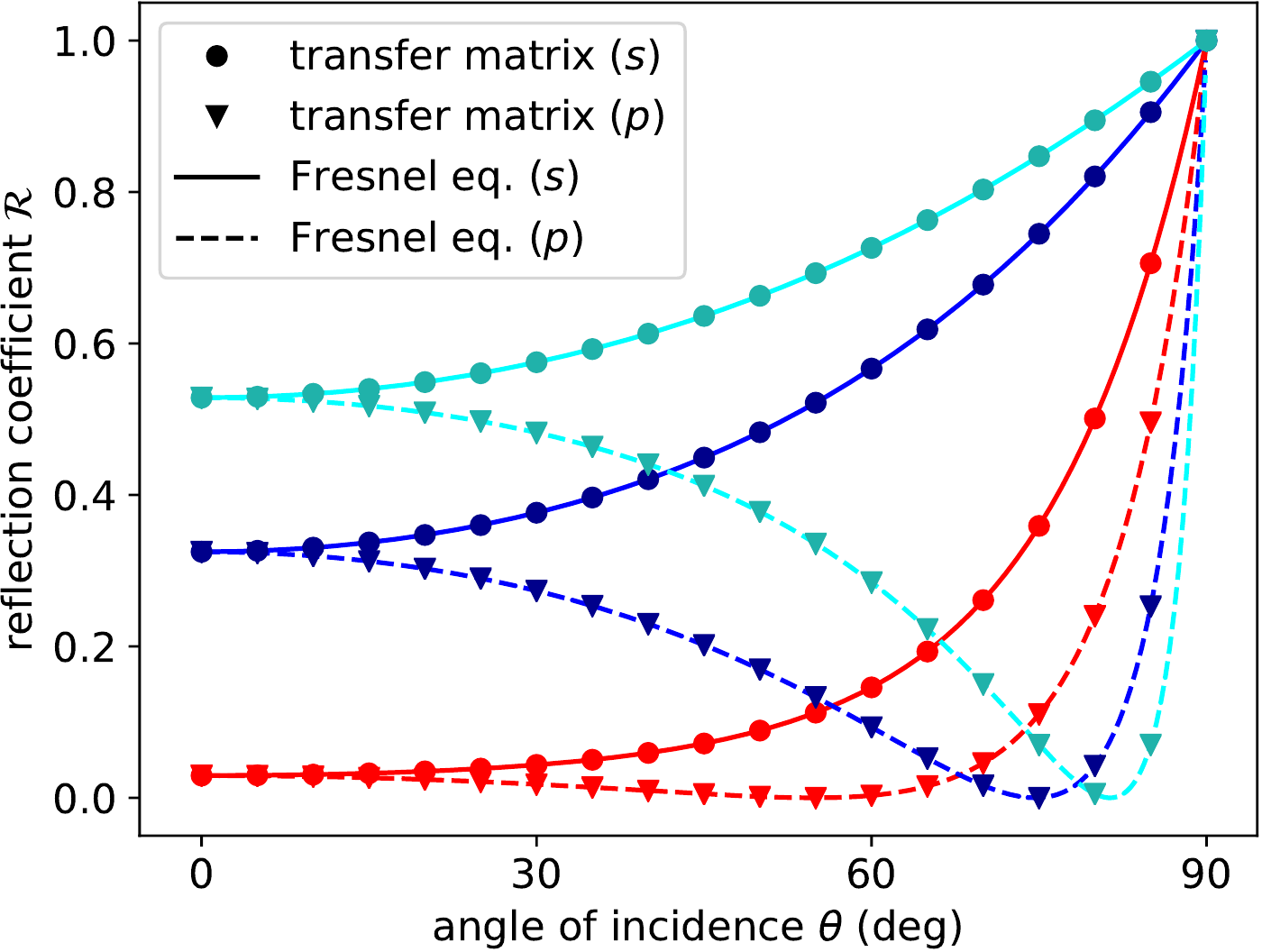}
    \caption{Angle-dependent reflectivity of an interface between media with optical dielectric constants $\varepsilon_1$ and $\varepsilon_2$. The red curves corresponds to optical dielectric constants of $\varepsilon_1$~=~1 and $\varepsilon_2$~=~2, the blue one to $\varepsilon_1$~=~3 and $\varepsilon_2$~=~40, and the turquoise one to $\varepsilon_1$~=~1, $\varepsilon_{2,\parallel}$~=~40, and $\varepsilon_{2,\perp}$~=~10.}
    \label{fresnel.fig}
\end{figure}

The reflection-related part of our implementation [Eqs.~\eqref{rS} and \eqref{rP}] can be compared to analytical formulas available for limiting cases. Since the transfer-matrix formalism does not entail any additional approximations, it has to reproduce these results identically. The first special case is a single interface between two dielectric, semi-infinite layers. The first layer has an optical dielectric constant $\varepsilon_\mathrm{1}$, the second one $\varepsilon_{\mathrm{2},\parallel}$ and $\varepsilon_{\mathrm{2},\perp}$. As shown in Fig.~\ref{fresnel.fig}, the transfer-matrix method yields the correct angle dependence of the reflectivity; the results are identical to those obtained from the generalized Fresnel expression, ${\cal R}^{(s/p)}(\theta) = |\zeta_-^{(s/p)}(\theta)/\zeta_+^{(s/p)}(\theta)|^2$, with
\begin{subequations}
\begin{align}
    \zeta_\pm^{(s)}(\theta) &= \varepsilon_\mathrm{1}\cos\theta\pm\varepsilon_\mathrm{1}^{1/2}\left(\varepsilon_{\mathrm{2},\parallel}-\varepsilon_\mathrm{1}\sin^2\theta\right)^{1/2}
\end{align}
and
\begin{align}
    \zeta_\pm^{(p)}(\theta) &= \varepsilon_{\mathrm{2},\parallel}\cos\theta\pm\xi_\mathrm{2}\varepsilon_\mathrm{1}^{1/2}\left(\varepsilon_{\mathrm{2},\perp}-\varepsilon_\mathrm{1}\sin^2\theta\right)^{1/2}.
\end{align}
\end{subequations}

As a second test case, we consider a dielectric mirror that consists of $m$ double layers of materials with optical dielectric constants $\varepsilon_1$ and $\varepsilon_2$. For a given wavelength $\lambda$, the reflectivity of this mirror is maximized for quarter-wave layers, \textit{i.e.} for thicknesses $d_1=\lambda/4\varepsilon_1^2$ and $d_2=\lambda/4\varepsilon_2^2$. In this case, the reflection coefficient for normal incidence is given as ${\cal R} = |\zeta_-/\zeta_+|^2$, with
\begin{align}
 \zeta_\pm = \varepsilon_{\mathrm{solv}}^2\varepsilon_2^{4m}\pm\varepsilon_N^2\varepsilon_1^{4m},
\end{align}
where $\varepsilon_{\mathrm{solv}}$ and $\varepsilon_N$ are the dielectric constants of the solvent and substrate above and below the stack, respectively.~\cite{dbr} Also in this case, analytic and matrix results yield equal values (not shown). These test cases demonstrate that the transfer-matrix formalism captures the dependence of the reflectivity on the angle of incidence, the anisotropy, and the thickness of the layers.

\subsection{Thiophene@MoS$_2$ on top of a dielectric mirror}\label{thiophene.sec}

\begin{figure}
    \centering
    \includegraphics[width=0.47\textwidth]{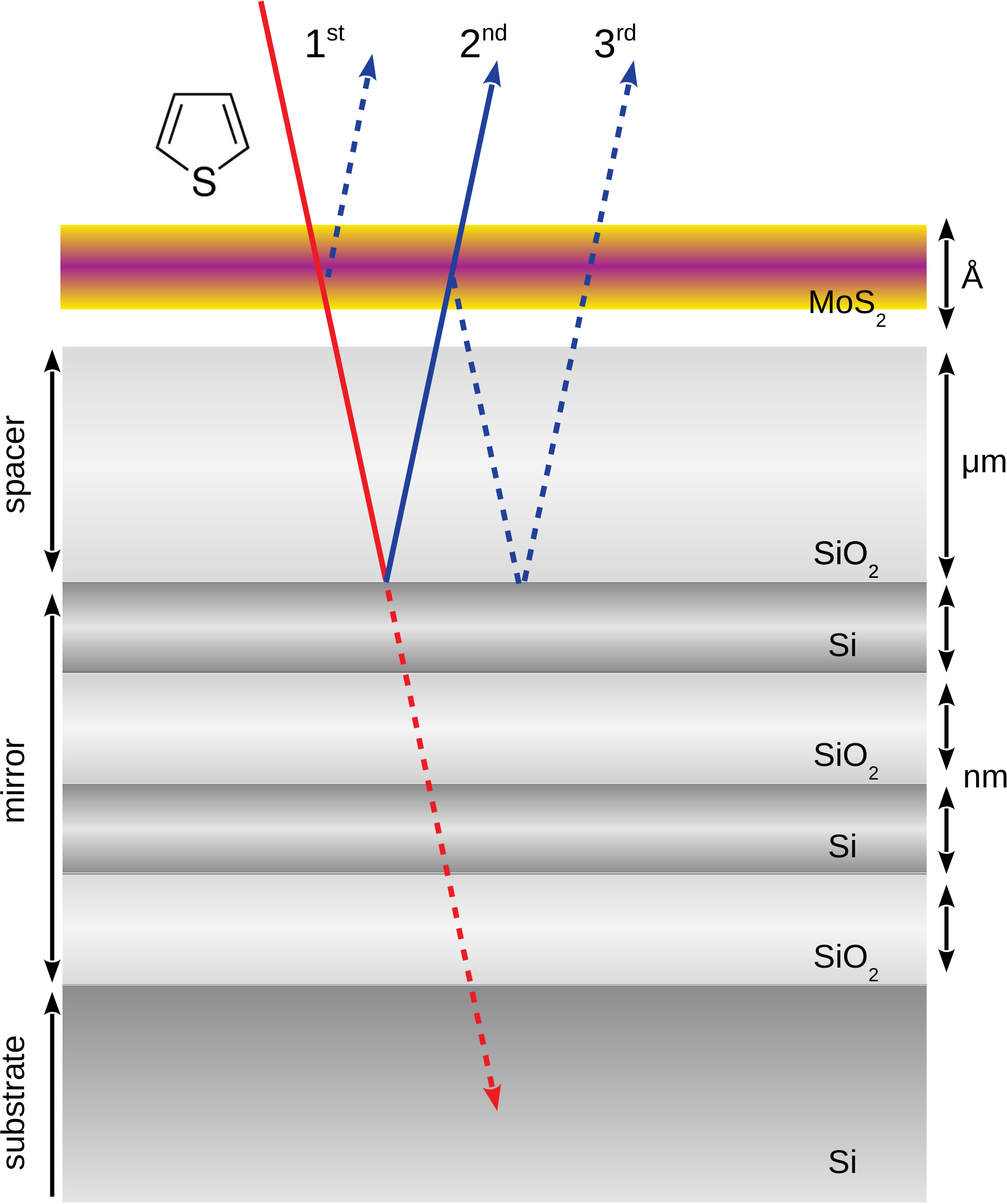}
    \caption{Schematic representation of the layered substrate used in the example. For the MoS$_2$ monolayer, we set\cite{laturia2018npj} $\varepsilon_\parallel$~=~15, $\varepsilon_\perp$~=~6 and $d$~=~4~\AA. It is separated by the next layer by 1~\AA\,\,of vacuum ($\varepsilon$~=~1). The spacer layer is made of SiO$_2$ ($\varepsilon$~=~2, $d$~=~1~$\mathrm{\mu}$m).~\cite{malitson1965josa} The two dual Si ($\varepsilon$~=~16, $d$~=~19.3~nm)\cite{aspnes1983prb} / SiO$_2$ ($d$~=~55.3~nm) layers underneath constitute a wavelength-selective mirror. The bulk substrate is of Si.}
    \label{parameters.fig}
\end{figure}

In this section, we test the whole formalism for a realistic system formed by a thiophene molecule adsorbed on a MoS$_2$ monolayer on top of a Si-based dielectric mirror. For convenience, we switch from atomic to conventional eV/\AA\,\,units. We consider the 1T at a distance 3~\AA\,away from the MoS$_2$ single-layer substrate atop the distributed Si/SiO$_2$ Bragg mirror separated by a large SiO$_2$ spacer layer (see Fig. \ref{parameters.fig}). The spacer layer has a depth on the order of microns, giving rise to notable delays of internally reflected fields. The thicknesses of the layers constituting the mirror are dictated by the wavelength that it is supposed to reflect, and thus on the nanometer scale. We do not use different values for the static and optical dielectric constants, since their values are fairly similar for the chosen materials close to the surface.\cite{laturia2018npj,malitson1965josa} This would not hold for (polar) solvents, which possess additional orientational degrees of freedom. 

We begin by examining the first ionization potential (IP) of 1T. In exact KS-DFT, this value is given as $-E_{\text{HOMO}}$.~\cite{perd1981prb} Within the adopted local-density approximation, we obtain values of $-E_{\text{HOMO}}$~=~6.10~eV and 6.13~eV \textit{in vacuo} and on top of the substrate, respectively. In other words, no relevant renormalization is obtained. However, if we estimate the IP by a $\Delta$SCF calculation, a different picture is observed: the IP decreases from 9.23~eV \textit{in vacuo} to 8.65~eV on the substrate. The absolute values are substantially higher than those of the KS levels, as expected, since the latter are always underestimated by the local-density approximation, as a consequence of self-interaction errors.~\cite{perdewZunger1981prb} More importantly, however, there is a sizeable reduction of the IP once the molecule is adsorbed on the substrate, which is a universal phenomenon. This shift is absent in the KS levels that are computed implicitly assuming a frozen substrate polarization, as no relaxation of the environment is taken into account. In the $\Delta$SCF calculation, on the other hand, the substrate is completely equilibrated with the ionized species. We note that the assumption of complete equilibrium is justified by the very similar values of the static and optical dielectric constants of the substrate; if remarkable differences were to exist, a non-equilibrium approach should be employed to allow only the fast environmental degrees of freedom to adapt to the ionization-induced charge-density variations.~\cite{pavel2008jpca,pavel2009jacs} Still, a significant renormalization of the energy levels would occur. This effect is not captured by $E_{\text{HOMO}}$ when simulating atomistically the combined system at the mere DFT level of theory with conventional exchange-correlation functionals.~\cite{liu2017jcp} A more sophisticated treatment of electron correlations, such as in the \textit{GW} approach,~\cite{garciaLastra2009prb, neaton2006prl, thygesen2009prl} is required for this purpose.

Proceeding with time-dependent simulations, we recall that an accurate model should take into account energy transfer between the molecule and the substrate. This can be expected to play a significant role in this case, since the excitation frequency of 1T overlaps with optical excitations of MoS$_2$ and SiO$_2$. A corresponding extension of the model, already developed for the case of pure solvent environments~\cite{ding2015jcp, corni2015jpca} and plasmonic nanoparticles,~\cite{pipoleCorni2016jpcc} will be the subject of future work; the results shown in this section  represent an initial survey of the problem. We apply to the system a normally incident, Gaussian-enveloped laser pulse
\begin{align}
    E_{\text{in}}(t) = E_0e^{-(t-t_\mu)^2/2t_\sigma^2}\cos(\omega_pt),
\end{align}
where $t_\mu=8$~fs and $t_\sigma=2$~fs;  the amplitude $E_0$ corresponds to a peak intensity of 1$\times$10$^{10}$~W/cm$^2$. The carrier frequency, $\omega_p$~=~5.6~eV, and the polarization vector, $\hat{\boldsymbol{\gamma}}$, are chosen to ensure resonance with the first bright excitation, corresponding to the transition from the highest occupied to the lowest unoccupied molecular orbital of 1T.~\cite{krumland2020jcp, cocchidraxl2015prb} The resonance frequency is not strongly affected by the presence of the substrate, implying that the band-gap renormalization does not translate into a reduced optical gap. This is quite generally the case; the electronic properties are more strongly influenced by polarizable environments than the optical properties. Energy levels are associated with charged systems, which evoke a strong polarization responses of the surrounding medium, as they create an attractive image charge that stabilizes the ion. The transition densities related to optical excitations, on the other hand, are overall charge-neutral and interact with the environment only via dipole-dipole and higher multipolar interactions. In the framework of MBPT, this corresponds to a reduced quasi-particle gap being compensated by the decrease of the exciton binding energy.

\begin{figure}
    \centering
    \includegraphics[width=0.47\textwidth]{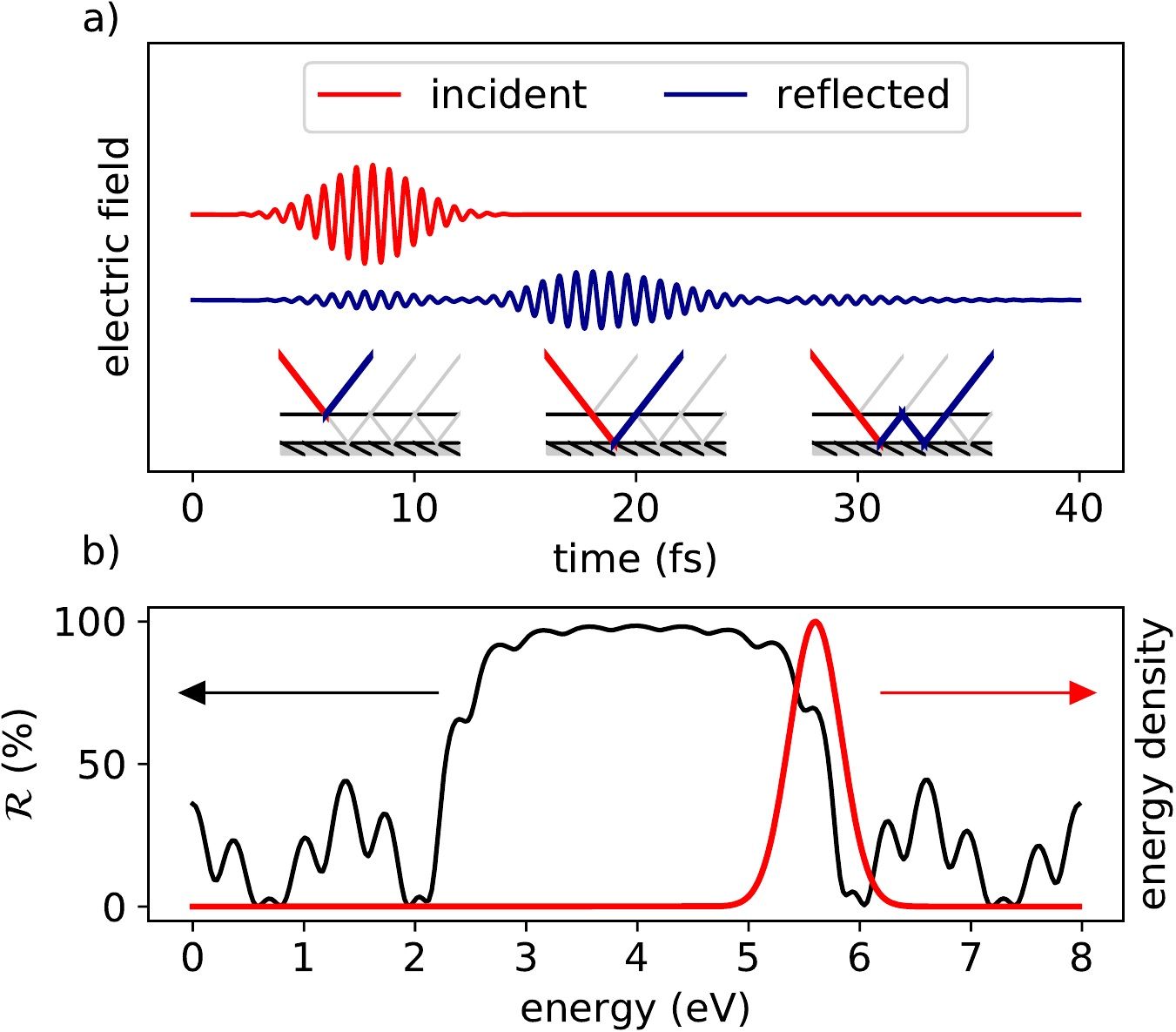}
    \caption{a) Incident and reflected fields in time domain. The sketches in the insets show the light pathways corresponding to the different reflected pulses. b) Reflectivity $\cal R$ (black) of the layered substrate characterized by the parameters given in Fig. \ref{parameters.fig} as well as the spectral energy density of the incident laser field (red).}
    \label{dbr.fjg}
\end{figure}

Before analyzing the induced electron dynamics, we consider the characteristics of the reflected field [see Fig.~\ref{dbr.fjg}a)]. Some of the radiation is immediately reflected at the interfaces close to the surface and interferes with the incident pulse. Due to the reflection at the deep-lying Bragg reflector, a strong delayed pulse leaves the substrate at $t=19$~fs. The shape of the pulse is altered with respect to the incident one, as clearly visible in its prolonged tail. This is a consequence of the strong frequency dependence of the reflectivity in the spectral region of the incident laser [Fig. \ref{dbr.fjg}b)]; the power spectrum is situated on the right flank of the main stopband of the Bragg mirror. A second field recurrence can be spotted around 30~fs. The corresponding wave does not exit the substrate after the reflection at the mirror, but is reflected at the surface, and reflected once more at the mirror before being transmitted through the surface. The combination of the incident field and the strong delayed pulses effectively turns probing the adsorbed molecule into a multi-pulse scenario. The delay between the pulses as well as their relative phase is related to the angle of incidence and the thickness of the spacer layer.
Modulating the substrate through the application of pressure to the substrate can be anticipated as an effective way to tune the phase relationship of the pulses.

\begin{figure}
    \centering
    \includegraphics[width=0.47\textwidth]{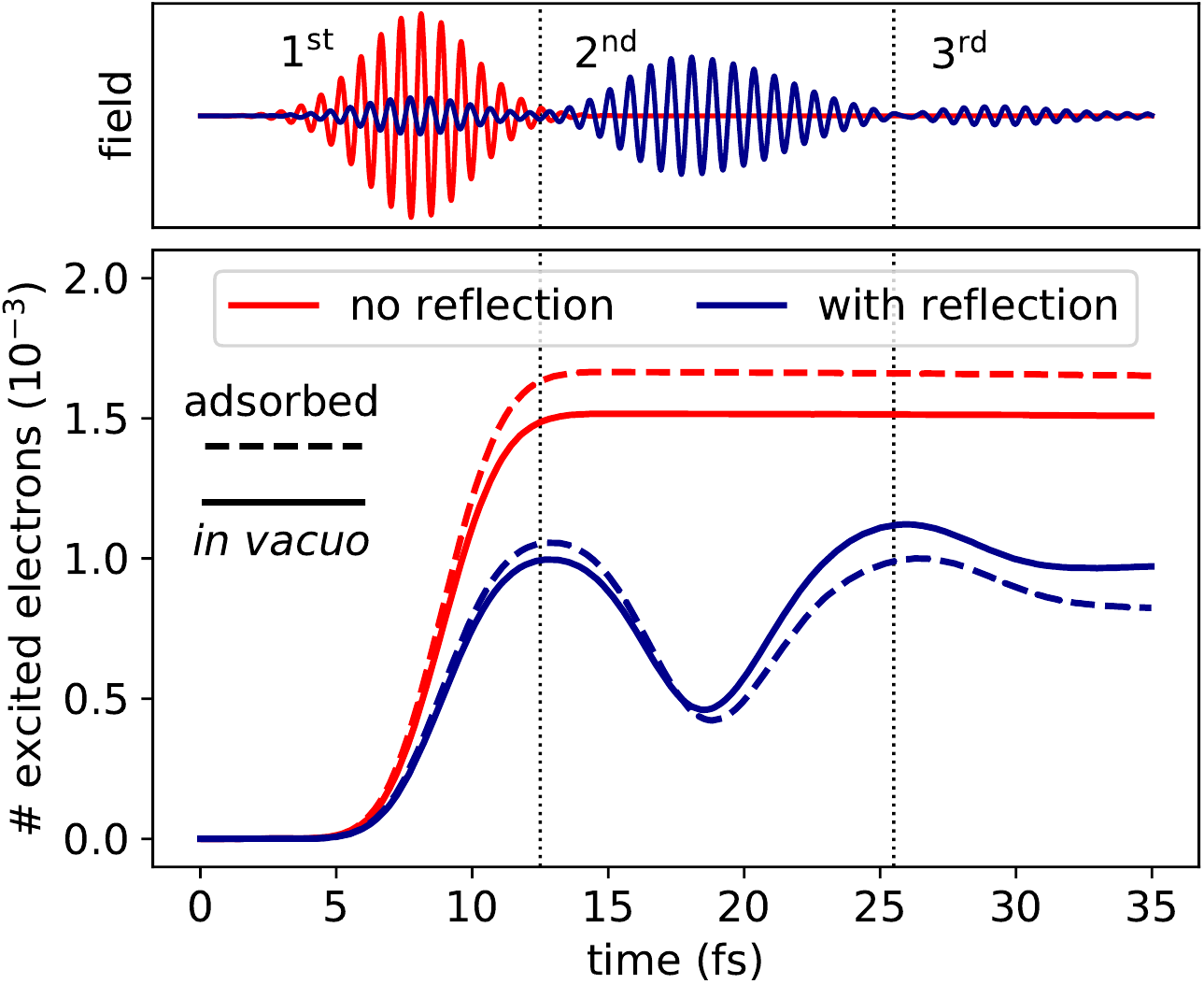}
    \caption{Excited-state population in laser-excited thiophene. The upper panel shows the incident (red) and reflected laser pulses (blue). The lower panel shows the number of excited electrons [Eq. \eqref{population.eq}] for four different cases; in the first two (red), the reflected field is not taken into, whereas in the other two (blue), it is included in the simulation. For the solid curves, the polarization response of the substrate is neglected, for the dashed ones, it is taken into account.}
    \label{populations.fig}
\end{figure}

We now turn to examining the electron dynamics induced by these fields. If the contribution of the reflected field is not included, the incident pulse leads to a level of excited-state population that remains constant after the pulse ends (solid red line in Fig.~\ref{populations.fig}). Taking into account the polarization response of the substrate, the achieved excited-state population increases (dashed red line in Fig. \ref{populations.fig}). Thus, the molecule/substrate coupling enhances the absorption strength. When the reflected field is included, the time evolution of the number of excited electrons becomes non-trivial. The initial increase is weaker, since the surface-reflected field destructively interferes with the incident one. The second pulse decreases the excited-state population to a minimum before raising it back to slightly above the initial level. This is an interference effect between the oscillating dipoles due to the incident and reflected fields, which have some arbitrary phase relationship.~\cite{coccia2019jcp} At the minimum around 18~fs, the induced dipole moment of the system is reduced to almost zero, while part of the excited-state population is retained. Afterwards, the dipole moment is built back up by absorption. The third pulse ultimately causes a small decrease of the excited-state population. This final decrease is equivalent to the one at the beginning of the second pulse, since the phase difference between the first and second pulse is the same as that between the second and the third. Notably, the final number of excited electrons is lower when the polarization response of the substrate is taken into account. This is a consequence of different phase relationships between the fields and the dipole moment, which can lead to either additional absorption or to deexcitation. The phase difference stems from the subtle polarization-induced shift of the transition frequency, which dictates the oscillation frequency of the dipole moment. 

\begin{figure*}
    \centering
    \includegraphics[width=\textwidth]{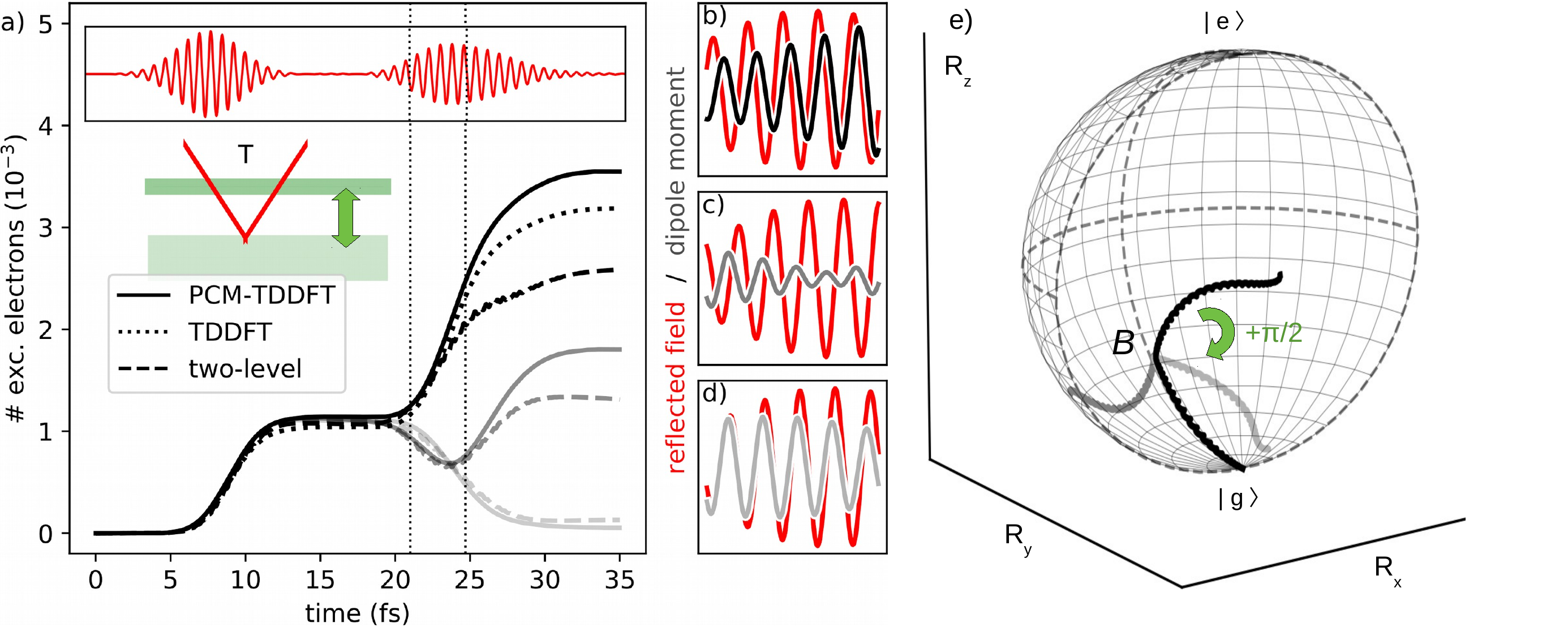}
    \caption{a) Number of excited electrons calculated from PCM/TDDFT (solid), TDDFT without reaction field (dotted, only black), and the two-level system (dashed). The different shades of grey correspond to different relative phases of the induced dipole moment and the reflected laser [panels b)-d)].
    These phase relationships correspond to spacer layer thicknesses of
    b) 1564.93~nm, c) 1623.61~nm, and
    d) 1604.05~nm
    respectively. The time window displayed in panels b)-d) is indicated in a) by the vertical dashed lines. e) Trajectories of the Bloch vector generated by the pulses with in panels b)-d) and calculated from the two-level system driven by scaled-up fields (see text). The same shades of grey are used for reference. The dashed lines in the sphere mark its intersections with the three Cartesian planes.}
    \label{phases.fig}
\end{figure*}

The phase relationship between the incident and reflected pulses can be tuned by changing the thickness of the spacer layer separating the surface and the Bragg mirror (see Fig. \ref{parameters.fig}). This can give rise to different interference effects that are reflected in the population dynamics.~\cite{giulia2020pccp} By deliberately choosing the layer thickness to have coherent incident and reflected fields, the reflected pulse further excites the molecule [see Fig.~\ref{phases.fig}a) and b)]. In this case, the reflected field is phase-shifted by $-\pi/2$ with respect to the dipole moment induced by the incident pulse. In the second case, the phase difference is $+\pi/2$ [Fig.~\ref{phases.fig}a) and c)]: This corresponds to the previously discussed scenario (Fig.~\ref{populations.fig}). The dipole moment is reduced to approximately zero, at which point it starts increasing again, as the reflected field has enforced the dipole-field phase relationship proper for absorption. The vanishing dipole moment at the minimum does not imply a stationary electronic system; the electron dynamics induced by the phase-shifted incident and reflected pulses overlap such that the dipole moments induced by such fields compensate each other at this point. A layer thickness bringing the dipole moment and the reflected field in phase with respect to each other leads to a depopulation of the excited state [Fig.~\ref{phases.fig}a) and d)]. At the end, the molecule is essentially back in its ground state.

In order to spot possible artifacts arising from adiabatic TDDFT,~\cite{maitra2016jcp} we compare the results of such calculations with those obtained from a two-level system, ruled by the exactly solvable Hamiltonian
\begin{align}\label{two_level.eq}
    \hat{\cal H}(t) = \Omega_{ge}|e\rangle\langle e|-\boldsymbol{\mu}_{ge}\cdot\textbf{E}(t)\big(| g\rangle\langle e|+|e\rangle\langle g|\big).
\end{align}
In Eq.~\eqref{two_level.eq}, the ground-state energy is shifted to zero, such that $\Omega_{ge}$ corresponds to the excitation energy from the ground state $|g\rangle$ to the excited state $|e\rangle$, with transition dipole moment $\boldsymbol{\mu}_{ge}$. The small static dipole moments of 1T in $|g\rangle$ and in $|e\rangle$ are neglected. Writing the total wavefunction as 
\begin{align}
    |\Psi(t)\rangle = c_g(t)|g\rangle + c_e(t)|e\rangle,
\end{align}
the time-dependent Schr\"odinger equation $i\partial/\partial t|\Psi(t)\rangle$ = $\hat{\cal H}(t)|\Psi(t)\rangle$ can be cast into matrix form and solved numerically by propagating in time the coefficients $c_g(t)$ and $c_e(t)$. The adopted system-related parameters, $\Omega_{ge}$~=~5.6~eV and $\boldsymbol{\mu}_{ge}$~=~1.86~D, result from a linear-response TDDFT calculation \textit{in vacuo} of the absorption spectrum of 1T. For consistency, the same electric field $\textbf{E}(t)$ is used in both RT-TDDFT and model calculations. The propagation of $c_g(t)$ and $c_e(t)$ is done with a fourth-order Runge-Kutta scheme.

We find qualitative agreement between the populations $|c_e(t)|^2$ of the excited state $|e\rangle$ and the number of excited electrons $N_e(t)$ calculated from TDDFT using Eq.~\eqref{population.eq},
 see Fig. \ref{phases.fig}d): in all three cases, the trend is the same. However, the final population predicted by TDDFT is notably lower compared to the model in cases of constructive interference [Fig. \ref{phases.fig}a)] and intermediate phase relationship [Fig. \ref{phases.fig}c)]. This, however, does not necessarily expose failures of adiabatic TDDFT;~\cite{krumland2020jcp} differences may as well be due to additional states playing a role in the dynamics. If this is the case, the reason for the quantitative mismatch lies in the insufficiency of the two-level system as a model for the molecule. Regardless, the qualitatively identical results obtained with respect to the model prove that adiabatic TDDFT can be a useful tool to explore the coherent dynamics of adsorbed molecules.~\cite{krumland2020jcp}

The achieved qualitative agreement between the results of the exact two-level model and TDDFT allows us to exploit this comparison also in the interpretation of the dynamics. Let us consider the Bloch vector, $\textbf{R}(t)$, in a frame rotating with the transition frequency $\Omega_{ge}$. This vector has components
\begin{subequations}
\begin{align}
R_x(t) &=  2\Re{\left\lbrace c_g^*(t)c_e(t)e^{-i\Omega_{ge}t}\right\rbrace}\\
R_y(t) &=  2\Im{\left\lbrace c_g^*(t)c_e(t)e^{-i\Omega_{ge}t}\right\rbrace}\\
R_z(t) &= |c_e(t)|^2-|c_g(t)|^2.
\end{align}
\end{subequations}
The absence of decoherence mechanisms in the adopted formalism implies that $\textbf{R}(t)$ always has unit length, \textit{i.e.} lies on the Bloch sphere depicted in Fig.~\ref{phases.fig}e). As the system is initially in its ground state, the trajectories start at the south pole of the sphere. During the first pulse, the three $\textbf{R}(t)$, corresponding to the different phases of the reflected pulse, congruently move up the side of the sphere towards the branching point $B$. This first path segment from the ground state to $B$ is within the $R_x$~=~0 plane, indicating resonance. The differences in the delayed reflected fields, however, lead to a splitting; $\textbf{R}(t)$ moves away from $B$ in different directions. The angles with which $\textbf{R}(t)$ emerges from $B$ are directly related to the phase difference between the first and second pulse. Trajectories are curved, indicating a time-dependent detuning, which is a consequence of the altered spectrum of the reflected pulse. The shape of the second segment starting at $B$ is essentially the same in all three cases, and they are just rotated versions of each other. This suggests that increasing the thickness of the spacer layer leads to a continuous rotation of the second segment around the branching point $B$ within the Bloch sphere, giving rise to the variable observed population and coherence dynamics. We notice that the presented line of reasoning holds exactly in the two level system. A quantitative analysis of this sort on RT-TDDFT calculations is not straightforward for the lack of the many-body wave function of the system, and thus goes beyond the scope of this work. However, we acknowledge that it may be potentially useful to reveal departure of the results from the two-state model.

All of the above-described phenomena take place within the linear regime. The trajectories in Fig.~\ref{phases.fig}e) are magnified: They have been obtained by multiplying all field strengths by a factor of 13. This scaling does not noticeably affect the shape of the curves, as the maximum excited-state population is still well below 50\% (marked by the equator of the sphere), where stimulated emission, \textit{i.e.} a nonlinear effect, starts to outweigh. Avoiding fields that are strong and/or long enough to cause population inversion and Rabi oscillations is imperative for obtaining reliable results with adiabatic RT-TDDFT.~\cite{krumland2020jcp,fuks2011pbr,raghunathan2011jctc,raghunathan2012jcp}

\section{Summary, conclusions, and outlook}\label{conclu.sec}
We have presented LayerPCM, an effective approach, based on the well-established polarizable continuum model combined with RT-TDDFT, to account for the electrostatic interactions between quantum-mechanically modelled molecules physisorbed on dielectric substrates. 
The key feature of the proposed method, which has been implemented in the \textsc{Octopus} code (v9.0), is to effectively model the dielectric properties of anisotropic layered substrates. 
It is therefore ideally applicable to low-dimensional materials and heterostructures, including complex patterns hosting adsorbed molecules.
After having tested the robustness of the implementation, we have shown that the LayerPCM can be successfully adopted to compute the polarization-induced renormalization of the frontier energy levels of the adsorbate molecules.
Furthermore, we have demonstrated its build-in suitability for laser-induced ultrafast dynamics simulations within RT-TDDFT.
In these calculations, we have explicitly accounted for the electric fields emerging from Fresnel-reflection at the substrate, which, according to the complexity of the layered surface, can assume non-trivial shapes and thus profoundly affect the electron dynamics in the molecule. Interference effects in the coherent dynamics of the adsorbate can emerge upon pronounced retardation effects of the field after laser irradiation.
The comparison of our numerical results with those obtained from an exactly solvable two-level model enables the quantitative analysis of the excited-state population, including the polarization effects induced by the substrate.

The numerical efficiency of the proposed approach offers new opportunities to simulate complex hybrid structures with nano-engineered substrates from first principles at affordable computational efforts. Coupling LayerPCM with DFT enables to capture the key effects of the substrate on the electronic structure of the adsorbed molecules, such as the energy level renormalization, without the need for atomistic simulation of the entire hybrid system. Also, it allows for a computationally efficient modulation of the substrate characteristics, such as the layer thickness, the dielectric function, and the anisotropy, in order to explore different scenarios. 
The advantages of LayerPCM in the time-domain are even more striking. The inclusion of reflected fields allows reproducing non-trivial physical effects that resemble those induced by multi-pulse spectroscopies. 

The presented implementation is the first step towards a comprehensive description of electromagnetic interactions between molecules and layered substrates, treated implicitly.  
Future steps entail the extension of LayerPCM to include frequency-dependent dielectric functions and the coupling with near fields.


\section*{Data Availability Statement}
The data that support the findings of this study are available from the corresponding author upon reasonable request.


\section*{Acknowledgements}
We are thankful to Prof. Roberto Cammi for useful discussions on some aspects of the theory. This work was funded by the Deutsche Forschungsgemeinschaft (DFG, German Research Foundation) - project number 182087777 - SFB 951, by the German Federal Ministry of Education and Research (Professorinnenprogramm III), and by the State of Lower Saxony (Professorinnen für Niedersachsen). Computational resources were provided by the North-German Supercomputing Alliance (HLRN), project bep00076. GG and SC  acknowledge funding from the ERC under the grant ERC-CoG-681285 TAME-Plasmons.

\appendix*

\section{Detailed derivation of the reflectivities}
Here, we derive the formulas for the reflectivity of the substrate in more detail. Inside a layer with dielectric tensor $\text{diag}\{\varepsilon_\parallel,\varepsilon_\parallel,\varepsilon_\perp\}$, Gauss' law $\nabla\cdot(\varepsilon\textbf{E})$~=~0 can be rearranged to read:
\begin{align}
 \nabla\cdot\textbf{E} = \frac{\varepsilon_\parallel-\varepsilon_\perp}{\varepsilon_\parallel}\frac{\partial E_z}{\partial z}.
\end{align}
Thus, we can write
\begin{align}
\nabla\times\nabla\times\textbf{E} &= -\nabla^2\textbf{E}+\nabla(\nabla\cdot\textbf{E})\nonumber\\
                                   &= -\nabla^2\textbf{E}+\frac{\varepsilon_\parallel-\varepsilon_\perp}{\varepsilon_\parallel}\nabla\frac{\partial E_z}{\partial z}.
\end{align}
Combining this result with Faraday's law, $\nabla\times\textbf{E} =-\partial{\textbf{B}}/\partial t$, and Ampere's law, $c^2\nabla\times\textbf{B} = \varepsilon\partial{\textbf{E}}/\partial t$, we obtain the wave equation for the electric field,
\begin{align}
    -\nabla^2\textbf{E}+\frac{\varepsilon_\parallel-\varepsilon_\perp}{\varepsilon_\parallel}\nabla\frac{\partial E_z}{\partial z} - \frac{1}{c^2}\varepsilon\frac{\partial^2\textbf{E}}{\partial t^2} = 0.
\end{align}
A Fourier transform with respect to both $\textbf{k}$ and $\omega$ yields
\begin{align}
k^2\textbf{E}+\frac{\varepsilon_\perp-\varepsilon_\parallel}{\varepsilon_\parallel}\textbf{k}k_zE_z + \left(\frac{\omega}{c}\right)^2\varepsilon\textbf{E}= 0.\label{waveFourier.eq}
\end{align}
Since the $k$-vector in the incident medium is assumed to be given, its in-plane component $k_\parallel$ is known throughout all the media, as it is maintained. The layer-specific out-of-plane components arise from the usual condition for non-trivial solutions of Eq. \eqref{waveFourier.eq}, \textit{i.e.} the determinant of the coefficient matrix must vanish. As the determinant is a polynomial of order 6, we obtain as many out-of-plane wavenumbers. For each of these modes, the solution of Eq. \eqref{waveFourier.eq} yields a corresponding polarization vector. However, the number of modes is reduced from 6 to 4 by Gauss' law, which poses additional constraints. The four wavevectors, corresponding to forward- and backward-propagating $s$- and $p$-polarized waves, are given in Eqs. \eqref{kS} and \eqref{kP}. They are associated with the polarization vector in Eqs. \eqref{pS} and \eqref{pP}. The two polarizations are independent and can be treated individually, as the in-plane/out-of-plane anisotropy does not couple them. Starting with the $s$~polarized field in layer $n$, we write down the electric field as a generic superposition of waves characterized by Eq. \eqref{kS} and \eqref{pS}:
\begin{subequations}
\begin{align}
    \textbf{E}_n^{(s)}(\textbf{r},t) = e^{i(k_\parallel x-\omega t)}&\left(E_{n,+}^{(s)}{\boldsymbol{\gamma}}^{(s)}_{n,+}e^{ik_{n,z}^{(s)}(z-z_{n-1})}\right.\nonumber\\&\left.+E_{n,-}^{(s)}{\boldsymbol{\gamma}}_{n,-}^{(s)}e^{-ik_{n,z}^{(s)}(z-z_{n-1})}\right),
\end{align}
where $E^{(s/p)}_{\pm}$ are unknown coefficients. The magnetic field associated with this electric field is obtained from Fourier-transformed Faraday's law, $\textbf{k}\times\textbf{E} = \omega\textbf{B}$, as
\begin{align}
    \textbf{B}_n^{(s)}(\textbf{r},t) = e^{i(k_\parallel x-\omega t)}&\left(E_{n,+}{\boldsymbol{\eta}}^{(s)}_{n,+}e^{ik_{n,z}^{(s)}(z-z_{n-1})}\right.\nonumber\\&\left.+E_{n,-}{\boldsymbol{\eta}}_{n,-}^{(s)}e^{-ik_{n,z}^{(s)}(z-z_{n-1})}\right),
\end{align}
\end{subequations}
where the polarization vectors associated with $\textbf{B}$ are given by
\begin{align}
 {\boldsymbol{\eta}}^{(s)}_\pm=\frac{\textbf{k}^{(s)}_\pm\times{\boldsymbol{\gamma}}^{(s)}_\pm}{\omega} = \left(\mp \frac{k_z^{(s)}}{\omega},0,\frac{k_\parallel}{\omega}\right).
\end{align}
The magnetic field $\textbf{B}$ is thus expressed in terms of the amplitudes of the electric fields, $E^{(s/p)}_{\pm}$, at the cost of the polarization vectors,  ${\boldsymbol{\eta}}^{(s)}_\pm$, being no longer unit vectors. Considering the boundary conditions $\delta\textbf{E}_\parallel=0$ and $\delta\textbf{B}_\parallel=0$ (the latter is valid for non-magnetic media) at the interface at $z_n$ between layer $n$ and $n+1$, we obtain
\begin{subequations}
\begin{align}
E_{n,+}^{(s)}e^{-ik_{n,z}^{(s)}d_n}+E_{n,-}^{(s)}e^{ik^{(s)}_{n,z}d_n} &= E_{n+1,+}^{(s)}+E_{n+1,-}^{(s)} \\
k^{(s)}_{n,z}\left(-E_{n,+}^{(s)}e^{-ik_{n,z}^{(s)}d_n}+E_{n,-}^{(s)}e^{ik^{(s)}_{n,z}d_n}\right) &= k^{(s)}_{n+1,z}\left(-E_{n+1,+}^{(s)}+E_{n+1,-}^{(s)}\right).
\end{align}
\end{subequations}
In matrix form, these equations read:
\begin{align}
 \begin{pmatrix}
E_{n,+}^{(s)}\\E_{n,-}^{(s)}
 \end{pmatrix}
 =P_nD_n^{-1}D_{n+1}
 \begin{pmatrix}
E_{n+1,+}^{(s)}\\E_{n+1,-}^{(s)}
 \end{pmatrix}, \label{constMatrixD.eq}
 \end{align}
 where 
 \begin{align}
 D_n &=
     \begin{pmatrix}
1 & 1 \\ -k_{n,z}^{(s)} & k_{n,z}^{(s)}
 \end{pmatrix}
\end{align}
and
\begin{align}
    P_n &= \text{diag}\{e^{ik_{n,z}^{(s)}d_n},e^{-ik_{n,z}^{(s)}d_n}\}.
\end{align}
After carrying out explicitly the matrix multiplications for all the inner layers, and sandwiching them between the first and final matrices
\begin{align}
    \begin{pmatrix}
    k_{s,z}^{(s)} & -1 \\
    k_{s,z}^{(s)} & 1
    \end{pmatrix}\hspace{0.3cm}\text{and}\hspace{0.3cm}
      \begin{pmatrix}
    1 & 1 \\
    -k_{N,z}^{(s)} & k_{N,z}^{(s)}
    \end{pmatrix},
\end{align}
we end up with Eqs. \eqref{ts} and \eqref{ds}.
In the case of $p$-polarized light, it is more convenient to start from the magnetic field
\begin{subequations}
    \begin{align}
    \textbf{B}_n^{(p)}(\textbf{r},t) = e^{i(k_\parallel x-\omega t)}&\left(B_{n,+}\hat{\boldsymbol{\eta}}^{(p)}_{n,+}e^{ik_{n,z}^{(p)}(z-z_{n-1})}\right.\nonumber\\&\left.+E_{n,-}\hat{\boldsymbol{\eta}}_{n,-}^{(p)}e^{-ik_{n,z}^{(p)}(z-z_{n-1})}\right)
\end{align}
rather than from the electric field. Thus, in this case, the magnetic-field polarization vectors $\hat{\boldsymbol{\eta}}^{(p)}_\pm$~=~$(0,1,0)$ are unit vectors, whereas those of the electric field, ${\boldsymbol{\gamma}}^{(p)}$, are not. The electric field is expressed in terms of the magnetic-field amplitudes,
\begin{align}
    \textbf{E}_n^{(s)}(\textbf{r},t) = e^{i(k_\parallel x-\omega t)}&\left(B_{n,+}^{(s)}{\boldsymbol{\gamma}}^{(s)}_{n,+}e^{ik_{n,z}^{(s)}(z-z_{n-1})}\right.\nonumber\\&\left.+B_{n,-}^{(s)}{\boldsymbol{\gamma}}_{n,-}^{(s)}e^{-ik_{n,z}^{(s)}(z-z_{n-1})}\right),
\end{align}
\end{subequations}
and, from $c^2\mathbf{k}\times\textbf{B}$~=~$-\omega\varepsilon\textbf{E}$, we have
\begin{align}
    {\boldsymbol{\gamma}}^{(p)}_\pm = -\frac{c^2\varepsilon^{-1}}{\omega}\textbf{k}^{(s)}\times\hat{\boldsymbol{\eta}}^{(p)}_\pm = \left(\pm \frac{c^2}{\omega\varepsilon_\parallel} k_z^{(p)}, 0, \frac{c^2}{\omega\varepsilon_\perp}k_\parallel\right).
\end{align}
Matching the in-plane components of adjacent layers according to $\delta\textbf{E}_\parallel=0$ and $\delta\textbf{B}_\parallel=0$ yields:
\begin{subequations}
\begin{align}
B_{n,+}^{(p)}e^{-ik_{n,z}^{(p)}d_n}+B_{n,-}^{(p)}e^{ik^{(p)}_{n,z}d_n} &= B_{n+1,+}^{(p)}+B_{n+1,-}^{(p)} \\
\frac{k^{(p)}_{n,z}}{\varepsilon_{n,\parallel}}\left(B_{n,+}^{(p)}e^{-ik_{n,z}^{(p)}d_n}-B_{n,-}^{(p)}e^{ik^{(p)}_{n,z}d_n}\right) &= \frac{k^{(s)}_{n+1,z}}{\varepsilon_{n+1,\parallel}}\left(B_{n+1,+}^{(s)}-B_{n+1,-}^{(s)}\right),
\end{align}
\end{subequations}
which, in matrix form, read:
\begin{align}
 \begin{pmatrix}
B_{n,+}^{(s)}\\B_{n,-}^{(s)}
 \end{pmatrix}
 =P_nD_n^{-1}D_{n+1}
 \begin{pmatrix}
B_{n+1,+}^{(s)}\\B_{n+1,-}^{(s)}
 \end{pmatrix}, 
 \end{align}
where
  \begin{align}
 D_n =
     \begin{pmatrix}
1 & 1 \\ -k^{(p)}_{n,z}/\varepsilon_{\parallel,n} & k^{(p)}_{n,z}/\varepsilon_{\parallel,n}
 \end{pmatrix}
 \end{align}
 and
 \begin{align}
 P_n = \text{diag}\{e^{ik_{n,z}^{(s)}d_n},e^{-ik_{n,z}^{(s)}d_n}\}.
\end{align}
The same steps as for $s$-polarization give rise to Eqs. \eqref{tp} and \eqref{dp}. Contrary to that case, however, the reflectivity in Eq. \eqref{rP} has to be supplied with a minus sign. The reason behind this is that we derived the magnetic-field related transfer matrix for $p$-polarization; whenever the magnetic field experiences a phase shift upon reflection at an interface, the corresponding electric field does not, and \textit{vice versa}. The minus sign accounts for this difference.



%

\end{document}